\documentclass{aa}

\usepackage{graphicx}
\usepackage{txfonts}
\usepackage{amssymb}
\usepackage{lscape}
\usepackage[dvips]{epsfig}
\usepackage{comment}

\begin{document}

\title{A grid of polarization models for \\ Rayleigh scattering planetary atmospheres\thanks{Full Table \ref{tab:cdstable} is only available in electronic form at the CDS via anonymous ftp to cdsarc.u-strasbg.fr (130.79.128.5) or via http://cdsweb.u-strasbg.fr/cgi-bin/qcat?J/A+A/504/259}
}

\author{E. Buenzli \& H.\,M. Schmid}

\offprints{Esther Buenzli, \email{ebuenzli@astro.phys.ethz.ch}}

\institute{Institute for Astronomy, ETH Zurich, 8093 Zurich, Switzerland}

\date{Received 30 January 2009 / Accepted 27 June 2009}

\abstract %context
{Reflected light from giant planets is polarized by scattering, offering the possibility of investigating 
atmospheric properties with polarimetry. Polarimetric measurements are available for the atmospheres of solar system planets, and instruments are being developed to detect and study the polarimetric properties of extrasolar planets.} 
%aim
{We investigate the intensity and polarization of reflected light from planets in a systematic way with a grid of model calculations. Comparison of the results with existing and future observations can be used to constrain parameters of planetary atmospheres.}
% method
{We present Monte Carlo simulations for planets with Rayleigh scattering atmospheres. We discuss the disk-integrated polarization for phase angles typical of extrasolar planet observations and for the limb polarization effect observable for solar system objects near opposition. The main parameters investigated are single scattering albedo, optical depth of the scattering layer, and albedo of an underlying Lambert surface for a homogeneous Rayleigh scattering atmosphere. We also investigate atmospheres with isotropic scattering and forward scattering aerosol particles, as well as models with two scattering layers.}  
% results
{The reflected intensity and polarization depend strongly on the phase angle, as well as on atmospheric properties, such as the presence of absorbers or aerosol particles, column density of Rayleigh scattering particles and cloud albedo. Most likely to be detected are planets that produce a strong polarization flux signal because of an optically thick Rayleigh scattering layer. Limb polarization depends on absorption in a different way than the polarization at large phase angles. It is especially sensitive to a vertical stratification of absorbers. From limb polarization measurements, one can set constraints on the polarization at large phase angles.}
% Conclusions
{The model grid provides a tool for extracting quantitative results from polarimetric measurements of planetary atmospheres, in particular on the scattering properties and stratification of particles in the highest atmosphere layers. Spectropolarimetry of solar system planets offers complementary information to spectroscopy and polarization flux colors can be used for a first characterization of exoplanet atmospheres.}


\maketitle  

\titlerunning{Polarization models for planetary atmospheres}   
\authorrunning{E. Buenzli \& H.M. Schmid}

\section{Introduction}
\label{sec:intro}
Light reflected from planetary atmospheres is generally polarized. 
The reflection is the result of different types of scattering particles with
characteristic polarization properties. Polarimetric observations
therefore provide information on the atmospheric structure and on 
the nature of scattering particles that complements other observations. 
Systematic model calculations are required to interpret the available polarimetry from solar system planets
and prepare for future polarimetric measurements of extrasolar planets. 

\paragraph{Scattering processes.}

Rayleigh scattering occurs on particles much smaller than the
wavelength of the scattered light. This process produces 100\%
polarization for a single right angle scattering.  
Rayleigh scattering is much stronger for short wavelengths because
the cross section behaves like $\sigma \propto 1/\lambda^4$, and it 
favors forward and backward scattering, which are both equally strong.
The blue sky in Earth's atmosphere is a well known example of 
Rayleigh scattering by molecules.

Aerosol haze particles with a size roughly comparable to the
wavelength can produce strongly forward directed
scatterings. Depending on the structure of the particle, a high
($p>90$\%) or low ($p\approx 20$\%) 
fractional polarization results for a scattering angle of $90^\circ$. For example, the maximum polarization
for scattering by optically thin zodiacal or cometary dust is not more
than $\approx 30$\% (e.g. Leinert et al.~\cite{leinert81}; Levasseur-Regourd et al.~\cite{levasseur96}), while a polarization close to 100\% is inferred for single scattering of haze
particles in Saturn's moon Titan (Tomasko et al.~\cite{tomasko08}).  

Liquid droplets in clouds produce a polarization because of 
refraction and reflection, which can be particularly high ($> 50$\%) for scattering angles of about $140^\circ$ for spherical water droplets,
corresponding to the primary rainbow (see e.g. Bailey
\cite{bailey07}). Clouds made of ice crystals reflect and refract light
in many different ways, and no distinct polarization features like rainbows are
expected, except locally, where ice crystals may have very similar structures. 

Multiple scatterings in planetary atmospheres randomize the polarization 
direction of the single scatterings and lower the 
observable polarization significantly. Therefore the net polarization of the
reflected light depends not only on the
scattering angle and the properties of the scattering particles, but
also on the atmospheric structure. For this reason it is not suprising
that a large diversity of polarization properties exists for the solar system planets.  

\paragraph{Observations.}

Venus shows a 
low ($<5$\%) negative polarization, which is a polarization parallel 
to the scattering plane, for most phase angles . In the blue and UV, a rainbow feature with a 
positive polarization of several percent is present 
(e.g. Coffeen \& Gehrels \cite{coffeen69}, Dollfus \& Coffeen
\cite{dollfus70}), indicating that the reflection occurs mainly from droplets in optically thick clouds
(Hansen \& Hovenier \cite{hansen74}). 
 
For the giant planets, only observations near opposition are possible
with earth-bound observations. Near opposition the disk-integrated
polarization is low because single back-scattering is unpolarized and
multiple scattering polarization cancels for a symmetric planet. 

With disk-resolved observations of Jupiter, Lyot (\cite{lyot29}) first detected that the Jovian poles show a strong limb polarization of order 5-10\%. 
To understand this effect one has to consider a back-scattering
situation at the limb of a sphere, where locally we have a configuration 
of grazing incidence and grazing emergence (for a plane parallel
atmosphere) for the incoming and the back-scattered photons, respectively.
Photons scattered upwards will mostly escape
without a second scattering, and photons scattered down have a low 
probability of being reflected towards us after the second 
scattering, but a high probability of being absorbed or undergoing
multiple scatterings. Thus photons that are reflected towards us
by two scatterings travel predominantely 
parallel to the surface. Because the polarization angle induced 
in a single dipole-type scattering process, like Rayleigh scattering, 
is perpendicular to the propagation direction of the
incoming photon, a polarization perpendicular to the limb is produced. 

Measurements at large phase angles $(\approx 90^\circ)$ for Jupiter with
spacecrafts detected a polarization of $\approx 50$\% for the poles while the 
polarization is much lower ($< 10$\%) for the equatorial region
(Smith \& Tomasko \cite{smith84}). 
The high polarization at the poles can be explained by reflection from 
a scattering aerosol haze layer, while the polarization at the equator is low
because of reflection from clouds. Towards short wavelengths (blue) the
polarization at the equator increases strongly, indicating that also
Rayleigh scattering contributes to the resulting polarization.

For Saturn the polarization is qualitatively similar to Jupiter with
an enhanced polarization at the poles at short wavelengths (blue).
In the red the polarization level of the poles is lower than for Jupiter
(Tomasko \& Doose \cite{tomasko84}). 

Uranus and Neptune display a strong limb polarization along the entire
limb (Schmid et al. \cite{schmid06a}; Joos \& Schmid
\cite{joos07}). Albedo spectra (e.g. Baines \& Bergstralh~\cite{baines86}) and the
polarization indicate that Rayleigh scattering is predominant 
in these atmospheres.

An interesting case is Saturn's moon Titan, which has a thick  
scattering layer of photochemical haze that produces a very
high disk-integrated polarization of $\sim 50~$\% in the B and
R band (Tomasko \& Smith \cite{tomasko82}). More recently the Huygens
probe measured the scattering and polarization properties of the aerosol
particles in great detail during its descent through Titan's
atmosphere (Tomasko et al. \cite{tomasko08}). 

The observations show that Rayleigh scattering is
an important polarigeneric process in atmospheres of solar system
objects, in particular for Uranus and Neptune, and for the equatorial regions of
Jupiter and Saturn. Besides Rayleigh scattering one has to consider the reflection from
haze particles (aerosols). Scattering by small aerosol particles
$(d<\lambda)$ may be approximated by Rayleigh scattering. For
large particles, $d \gtrsim \lambda$, the strong forward scattering effect
and the reduced polarization for right angle scattering cause
significant differences when compared to Rayleigh scattering. 

Clouds dominate in the atmosphere of Venus, and at 
longer wavelengths (red) also in Saturn and Jupiter. The reflection from clouds produces
only a low positive or even negative polarization signal in Venus,
Saturn or Jupiter, typically at a level $p <5~$\%. 
In a first approximation one may therefore 
treat clouds like a diffusely scattering layer producing no polarization.  

Polarimetric measurements of stellar systems with known extrasolar planets were attempted, but up to now
no convincing detection of the polarized reflected light from an
extrasolar planet has been made (Lucas et al.~\cite{lucas09}, 
Wiktorowicz~\cite{wiktorowicz09}). The deduced upper limits on the polarization
flux from the close-in planet indicate that these objects are
not covered with a well reflecting Rayleigh scattering layer.

\paragraph{Model calculations.}
  
The classical theory for the analytic solution of the multiple
scattering problem is treated in the 
seminal work of Chandrasekhar (\cite{chandrasekhar50}), from which the
polarization of conservative (non-absorbing) Rayleigh scattering
planets can be derived. Van de Hulst (\cite{vandehulst80}) gives a comprehensive overview on
theoretical work up to that time including many numerical model results. 

Schmid et al.~(\cite{schmid06a}) put together available model 
results useful for parameter studies of the polarization from Rayleigh scattering atmospheres.
This includes the following model results:
\begin{itemize}
\item{} Phase curves for the disk-integrated intensity and polarization for finite, 
        conservative (no absorption) Rayleigh scattering atmospheres for 
        different optical thicknesses and ground albedos from Kattawar \& Adams 
        (\cite{kattawar71}),
\item{} the limb polarization at opposition for semi-infinite Rayleigh scattering 
        atmospheres with different single scattering albedos
        derived from formulas and tabulated functions given
        in Abhyankar \& Fymat (\cite{abhyankar70}, \cite{abhyankar71}) and Chandrasekhar (\cite{chandrasekhar50}), 
\item{} the limb polarization at opposition for finite, conservative (no absorption) 
        Rayleigh scattering atmospheres for different optical thicknesses
        and ground albedos from tabulations given in Coulson et al. (\cite{coulson60}).
\end{itemize}

For Venus detailed models for the reflection from clouds were
developed, which demonstrate nicely the diagnostic potential of polarimetric measurements 
(e.g. Hansen \& Hovenier \cite{hansen74}). 
More recent modeling of the polarization from planets was performed
mainly to analyze and reproduce polarimetric observations of Jupiter
and Titan from spacecrafts
(e.g. Smith \& Tomasko \cite{smith84}; Braak et al. \cite{braak02};
Tomasko et al.~\cite{tomasko08}).  

Another line of investigation now concentrates on the 
expected polarization of extrasolar planets. The Rayleigh and Mie scattering polarization of close-in planets was
investigated by Seager et al. (\cite{seager00}). These calculations consider planets which are unresolved from their central star and the polarization signal is strongly diluted by the unpolarized stellar light. 

Stam et al.~(\cite{stam04}) modeled the polarization of a Jupiter-like extrasolar planet with methane absorption bands for three special cases and presented polarization spectra and wavelength integrated phase curves. Also monochromatic phase curves for a non-absorbing clear and a hazy atmosphere are available (Stam et al.~\cite{stam06}). Other studies determined the expected polarization from clouds of terrestrial planets (e.g. Bailey
\cite{bailey07}) or the polarization of extrasolar analogs to Earth (Stam \cite{stam08}).

Despite all these models systematic model
calculations are sparse in the literature. For finite Rayleigh scattering atmospheres, polarization phase curves have been calculated only for few selected cases. No results are available for the limb polarization of atmospheres with finite thickness and absorption.

It is the goal of this paper to present a grid of model results 
for Rayleigh scattering models with absorption and to explore the 
model parameter space in a systematic way. The results should allow a comparison
with observations and provide a tool for their interpretation. Additionally effects
of selected deviations from simple Rayleigh scattering models will be discussed. 
  
In the next section the paper describes our scattering model and the
Monte Carlo simulations. Section~\ref{sec:homo} presents the results from a
comprehensive Rayleigh scattering model grid covering the three atmosphere
parameters: single scattering albedo $\omega$, optical thickness of the Rayleigh
scattering layer $\tau_{\mathrm{sc}}$, and albedo of the underlying reflecting
surface $A_\mathrm{S}$. In Sect.~\ref{sec:beyond} we explore the effects of a mixure of isotropic and Rayleigh scattering, of particles with a forward scattering phase function, and of two polarizing layers. In Sect.~\ref{sec:wavelength} we discuss spectral dependences. Section~\ref{sec:special} highlights some special cases and diagnostic diagrams which may be of particular interest for the interpretation of
observational data. A discussion and conclusions are given in the final
section. Appendix \ref{sec:cdstables} describes the tables with the numerical results of our calculations of intensity and polarization phase curves for a grid of 333 model parameter combinations. These are available in electronic form at the CDS.

\section{Model description}
\label{sec:model}

Our planet model consists of a spherical body of radius $R$, illuminated 
by a parallel beam. This geometry is appropriate for not rapidely rotating 
planets with a large separation, $d\gg R$, from the parent star.    
Each surface element is approximated by a plane parallel atmosphere. This 
simplification is reasonable for planets 
without an extended, tenuous atmosphere. 

\subsection{Intensity and polarization parameters} \label{sec:polpar}

The intensity and polarization of the reflected light is described by the
Stokes vector $\mathbf{I} = (I,Q,U,V)$. The linear polarized
intensity or polarization flux is defined by the parameters 
$Q=I_0-I_{90}$ and $U=I_{45}-I_{135}$, where the indices
stand for the polarization direction with respect to a specified direction
in the selected coordinate system. In this paper only processes producing
linear polarization are studied and therefore the Stokes parameter $V$ 
for the circular polarization is omitted. 
We express the fractional polarization by the symbols
\begin{equation}
q = {Q \over I}, \; u = {U \over I}, \; p = {\sqrt{Q^2+U^2} \over I}\,,
\end{equation}

\noindent and the polarized intensity
\begin{equation}
p \cdot I = \sqrt{Q^2+U^2}\,.
\end{equation} 

For the study of the limb polarization in resolved solar system
planets at opposition we introduce the radial Stokes parameter $Q_r$, which is positive for an 
orientation of the polarization parallel to the radius vector $\vec{r}$ 
(perpendicular to the limb) and negative for an orientation
perpendicular to $\vec{r}$. The Stokes $U_r$ parameter is the 
polarization direction $\pm 45^\circ$ to the radius vector 
(see e.g. Schmid et al.~\cite{schmid06a} for an illustrative 
description of the radial polarization). The polarization fraction is 
represented by $q_r$ and $u_r$. 

The radial polarization curves $q_r(r)$ and $Q_r(r)$ can only be observed if the planetary disk is
well resolved. The measured radial profile depends strongly 
on the achieved spatial resolution. Because of the limited spatial resolution of
most observations it is very hard to exactly measure 
the polarization near the limb. It is much less difficult to
evaluate a disk-integrated polarization or polarization flux and to
estimate and correct the degradation of the observed value 
with respect to the intrinsic value with a simulation of the observational
resolution or point spread function. This approach is described in
detail in Schmid et al.~(\cite{schmid06a}) for seeing limited
polarimetry of Uranus and Neptune. 

Therefore we mainly discuss the intensity weighted
polarization $\langle q_r \rangle = \langle Q_r\rangle/I$, which is the equivalent to 
the disk-integrated radial polarization 
$\int Q_r(r) 2\pi\, r \,{\mathrm{d}}r $ normalized to the geometric albedo. 
The geometric albedo $A_{\mathrm{g}}$ is the disk-integrated reflected intensity 
of a given model at opposition normalized to the reflection of a white Lambertian disk. It corresponds to 
$I(0^{\circ})$ in our calculations.

The radial polarization curves are qualitatively similar for most models. The
shape of the intensity curve varies significantly from limb darkening to limb brightening for different
model parameteres and cannot
solely be described by the geometric albedo. Additionally we choose the Minnaert law exponent $k$ as fit parameter for the shape of the center-to-limb intensity curve. 
The Minnaert law for opposition is $I(r) = I_{\mathrm{r=0}} \mu^{2k-1}$, where
$\mu(r)=(1-r^2)^{1/2}$. This yields the following one-parameter fit curve 
$I(r) = I_{\mathrm{r=0}} (1-r^2)^{k-1/2}$. 

\subsection{Atmosphere parameters}
\label{sec:atmosphere}

The plane parallel atmosphere is assumed to consist of a homogeneous scattering 
layer that is either semi-infinite or finite with a reflecting (cloud or ground)
Lambertian surface layer with a surface albedo. The basic model atmospheres
are described by three parameters:  
\begin{itemize}
\item{} the single scattering albedo $\omega$, 
\item{} the (vertical) optical thickness for scattering, $\tau_{\mathrm{sc}}$, of the scattering layer,
\item{} the albedo $A_\mathrm{S}$ of the surface below the scattering layer.  
\end{itemize}

The single scattering albedo $\omega$ is defined by the
ratio between the scattering cross section $\sigma$ and the sum of absorption cross section $\kappa$
and scattering cross section $\sigma$, with the cross sections multiplied by the fractions 
of scattering or absorbing particles ($f_{\mathrm{sc}}$ or  $f_{\mathrm{abs}}$)
\begin{equation}
\omega = {f_{\mathrm{sc}}\sigma \over f_{\mathrm{abs}}\kappa + f_{\mathrm{sc}} \sigma}\,.
\end{equation}
The value $\omega=1$ indicates pure scattering (no absorption) while $\omega=0$ is
the other extreme of no scattering and just absorption (e.g. black dust).
Similarly, a surface albedo of $A_\mathrm{S}=0$ 
corresponds to a black surface, while a perfectly white Lambertian 
surface is defined by $A_\mathrm{S} = 1$.

The optical depth for scattering $\tau_{\mathrm{sc}}$ follows from the column density $Z$ of the scattering layer: 
$\tau_{\mathrm{sc}} =  Z \cdot \sigma$, where $\sigma$ is the scattering cross section per particle. 
The semi-infinite case corresponds to $\tau_{\mathrm{sc}}=\infty$.
We treat absorption like an addition of absorption optical depth to a
layer with a given scattering optical depth $\tau_{\mathrm{sc}}$, which is equivalent to reducing the single scattering albedo. This approach is suited for discussing the reflected
intensity and polarization inside and outside of absorption features
like CH$_4$ or H$_2$O-bands, where $\kappa$ differs dramatically while $\sigma$ is essentially equal.
Then the total optical thickness $\tau$ of the layer including absorption $\kappa$ is given by 
\begin{equation}
\tau = (f_{\mathrm{sc}}\sigma + f_{\mathrm{abs}}\kappa)\cdot Z \ = {\tau_{\mathrm{sc}} \over \omega}\,.
\end{equation} 

The basic model grid (Sect.~3) considers only Rayleigh scattering  
($\sigma=\sigma_{\mathrm{Ray}}$, $\tau_{\mathrm{sc}}=\tau_{\mathrm{Ray}}$) as scattering process and Lambert surfaces
with an albedo $A_\mathrm{S}$ below the scattering layer. Extensions, such as including non-polarizing isotropic scattering where $\sigma=\sigma_{\mathrm{Ray}} + \sigma_{\mathrm{iso}}$, haze layers or more than one scattering layer are discussed in Sect.~4.

\subsection{Geometric parameters}

\begin{figure}
         \resizebox{\hsize}{!}{\includegraphics{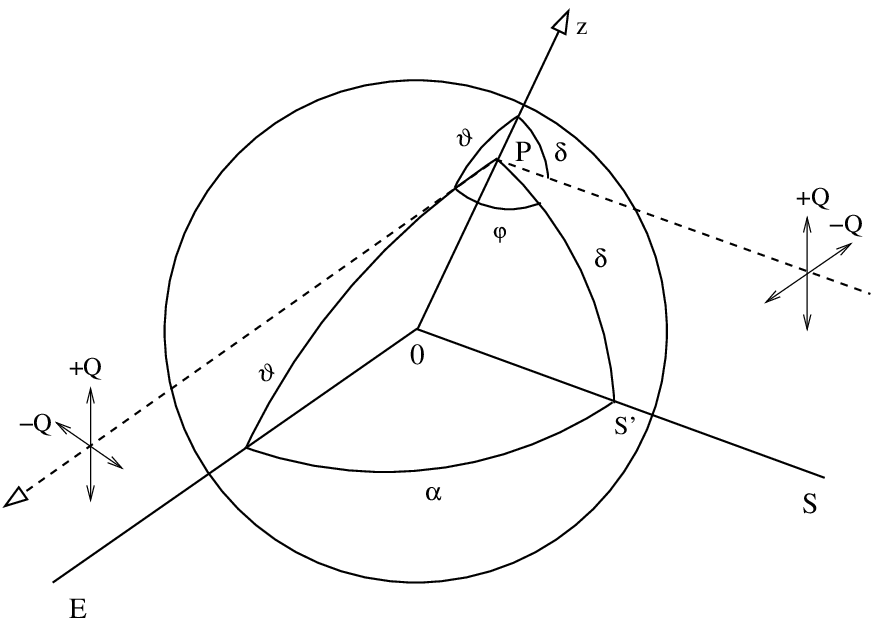}}
          \caption{Model geometry. The dashed line represents the trajectory of a reflected photon.}
          \label{fig:geometry} 
\end{figure}

The geometric parameters describe the location of the considered
surface point $P$ and the escape direction of the photons (Fig.~\ref{fig:geometry}).
A global coordinate system describes the orientation of the planet with respect to the star. Its polar axis is the surface normal at the sub-stellar point $S'$, and the location of each point $P$ is described by polar angle $\delta$ and azimuthal angle $\theta$ (not drawn). $\delta$ is also the photon's angle of incidence at point $P$. The escape direction, i.e. the location of the observer, is given by a polar angle $\alpha$ and azimuthal angle $\chi$ (not drawn). $\alpha$ is equivalent to the phase angle defined by the three (central) points: star or sun $S$, planet $0$, and observer $E$ (Earth). 

For the description of the scattering processes, a local coordinate system is set up at point $P$ for the plane parallel atmosphere with surface normal $z$ perpendicular to the planet surface in $P$, polar angle $\vartheta$ and azimuthal angle $\varphi$. 

In general, each point $P$ can have individual atmospheric properties. Then
the model outputs, the Stokes vector components $I$, $Q$ and $U$, depend each on seven parameters:

\begin{displaymath}
\mathbf{I}(\delta,\theta,\alpha,\chi,\tau_{\mathrm{sc}}(\delta,\theta),\,\omega(\delta,\theta),\,A_\mathrm{S}(\delta,\theta))\,.
\end{displaymath}

This description allows calculation of the reflected intensity and polarization
of each surface point on the illuminated hemisphere viewed from any direction. 
Obviously this large parameter space needs to be simplified
for a first parameter analysis. If we adopt the same atmospheric structure everywhere on the planet,
 $\tau_{\mathrm{sc}}, \omega$ and $A_\mathrm{S}$ are no longer functions of $\delta$ and $\theta$ and 
we obtain a rotationally symmetric model geometry with respect to 
the line $S-0$, which is independent of the azimuthal angle $\chi$. 

For extrasolar planets it will not be
possible to resolve the disk in the near future. For disk-integrated results we can eliminate the dependence  
of the reflected intensity and polarization on the surface point parameters $\delta$ and $\vartheta$. 
Because of the rotational symmetry of the geometric model, the
intensity and the polarization then only depend on the polar 
viewing angle or phase angle $\alpha$. Moreover the orientation of the
polarization signal is either parallel or perpendicular to the scattering
plane (the plane S-0-E), which we call the $Q$
polarization direction. $Q$ is defined positive for a polarization
perpendicular to the plane S-0-E and negative for a polarization parallel to this
plane. The $U$-polarization is zero in this coordinate system for
symmetry reasons.   

For the full disk the integrated intensity and polarization signals from a planet
depend on the following parameters:
\begin{displaymath}
% I(\alpha,\,\tau_{\mathrm Ray},\,\omega_{\mathrm Ray},\,\omega_{\mathrm Iso},\,A_{\mathrm S})\,, \\
% Q(\alpha,\,\tau_{\mathrm Ray},\,\omega_{\mathrm Ray},\,\omega_{\mathrm Iso}\,\,A_{\mathrm S})\,.
I(\alpha,\,\tau_{\mathrm{sc}},\,\omega,\,A_\mathrm{S})\,, \\
Q(\alpha,\,\tau_{\mathrm{sc}},\,\omega,\,A_\mathrm{S})\,.
\end{displaymath} 

For solar system planets at opposition we obtain a rotationally symmetric scattering geometry (viewing
direction is identical to the axis of symmetry of the geometric model). 
We then have a scattering model which depends only on $\delta$ or 
the normalized projected radius 
$r=\sin \delta$, and which is independent of $\theta$. The resulting 
polarization will be in the radial direction either parallel or 
perpendicular to the radius vector $\vec{r}$ and therefore our model output is
the radial Stokes parameter $Q_r$ (cf. Sect.~\ref{sec:polpar}). The Stokes $U_r$ parameter again
has to be zero for a spherically symmetric planet.

For exact opposition the dependences of the scattering model results 
can be described by the following parameters:
\begin{displaymath}
% I(r,\,\tau_{\mathrm Ray},\,\omega_{\mathrm Ray},\,\omega_{\mathrm Iso},\,A_{\mathrm S})\,, \\
% Q_r(r,\,\tau_{\mathrm Ray},\,\omega_{\mathrm Ray},\,\omega_{\mathrm Iso},\,A_{\mathrm S})\,.
I(r,\,\tau_{\mathrm{sc}},\omega,\,A_\mathrm{S})\,, \\
Q_r(r,\,\tau_{\mathrm{sc}},\omega,\,A_\mathrm{S})\,.
\end{displaymath} 
These are the center to limb intensity curve and the center to limb radial
polarization curves which both depend only on the atmospheric parameters.

\subsection{Monte Carlo simulations}
\label{sec:monte}

For our simulations we used the Monte Carlo code described in Schmid
(\cite{schmid92}), which was slightly adapted for the case of light reflection 
from a planet. Basically the code calculates  the random 
walk histories of many photons in the planet model atmosphere until the photons have 
escaped or are destroyed by an absorption process. After a sufficiently large
number have escaped, the scattering intensity and polarization of the
reflected light can be established for different lines of sight. 
In our calculations we assume that despite multiple scatterings the escaping
photons emerge at the same point where they penetrated into the planet. 
In each scattering process the photon undergoes a
direction and polarization change calculated from the appropriate phase matrix.
The linear polarization of the photons in the simulations is defined  by the 
orientation $\gamma$ of the electric vector for the photon's 
electromagnetic wave. In a given coordinate system we can then
evaluate the contribution to the Stokes intensity for each photon in $Q \propto
\cos 2\gamma$ and $U \propto \sin 2\gamma$ direction.  

The escaping photons have to be collected in discret 
direction bins (in our models phase angles $\alpha =2.5^\circ,7.5^\circ, \ldots$ with a finite bin width $\Delta\alpha=5^\circ$) to evaluate $I(\alpha)$ and $Q(\alpha)$. These are then a mean photon intensity and polarization for that bin. 
$\Delta\alpha$ should be small to resolve any structure in $I(\alpha)$ and $Q(\alpha)$,
but also sufficiently large to collect enough photons for results with small statistical
errors. The aim of our simulations is to reach at least the expected
precision of observational data. The rotational
symmetry imposed on our models helps to increase the bin size for the phase curve
interval $\alpha_k$, which behaves like $\alpha_k \propto \sin \alpha$. 
This means that we have to divide the photon count per bin 
by the factor $2\pi\sin\alpha\Delta\alpha$. The intensity is obtained by normalizing with the reflectivity of a white Lambertian disk.
For a given simulation the relative statistical
errors (photon shot noise) are particularly good for $\alpha \approx 90^\circ$, much less favorable for 
$\alpha=2.5^\circ$ and very bad for $\alpha = 177.5^\circ$ where only a few photons
will be collected, because the irradiated hemisphere of the planet 
is almost invisible for this phase angle. For the center-to-limb  curves we bin uniformly in $\delta=\arcsin(r)$ with a bin size of $\Delta\delta=5^\circ$, which requires an additional normalization by $2\pi\sin\delta\cos\delta\Delta\delta$. 

The number of photons per model was chosen such that the 
number of reflected photons in phase angle bins relevant for observations ($\alpha \approx 30^\circ - 120^\circ$) are about $N \approx 2 \cdot 10^6$ when integrated over the whole disk. This corresponds to an error in polarization $\Delta p = \sqrt{2/N} =0.1$\%. For the radial curves the total number of photons was increased such that the same precision was reached in most radial bins. No photons emerge at the exact phase angle $\alpha=0^\circ$. Therefore for the limb polarization calculations we count all photons that are in the bin $0^\circ < \alpha < 5^\circ$, even though the calculation for the radial polarization includes the assumption that $\alpha=0^\circ$. The error induced by this measure is smaller than the statistical error. %checked by making bins smaller, with only 0 to 2.5 the difference is well within the statistical error

A general guideline for the Monte Carlo technique for random walk problems 
is given in Cashwell \& Everett (\cite{cashwell59}) and many Monte Carlo
simulations for the investigation of light scattering are described in the
astronomical literature (see e.g. Witt~\cite{witt77}, Code \& Whitney~\cite{code95}, Wolf et al.~\cite{wolf99}).
In Schmid (\cite{schmid92}) a detailed description on many aspects of 
the employed Monte Carlo code are given; e.g. the general scheme of
the code, the required transformations between the involved coordinate
systems (star - planet, planet - plane parallel atmosphere, atmosphere
- photon), the determination of the free path
length, the treatment of isotropic scattering and Rayleigh scattering 
according to the Rayleigh phase matrix, an assessment of statistical errors, and a comparison with analytical calculations.

\section{Model results for a homogeneous Rayleigh scattering atmosphere}
\label{sec:homo}

This section discusses the model grid results for simple homogeneous Rayleigh scattering atmospheres described by parameters $\omega$, $\tau_{\mathrm{sc}}$ and $A_\mathrm{S}$ (cf. sec. \ref{sec:atmosphere}) We discuss phase curves (Sect.~\ref{sec:phcurve}) and radial profiles (Sect.~\ref{sec:opposition}) for selected cases and explore the full parameter space for disk-integrated results at $\alpha = 90^\circ$ and $\alpha = 0^\circ$ (Sect.~\ref{sec:param}).

Many of the general dependences of these model results on atmospheric parameters were already discussed in previous studies mentioned in the introduction (Sect.~\ref{sec:intro}). Compared to these our calculations are much more comprehensive and the extensive model grid results are provided in electronic form (see Appendix~\ref{sec:cdstables}). An overview of the dependence of observable quantities, such as intensity, fractional polarization and polarized intensity, on atmosphere parameters is presented in diagrams which may be useful for the interpretation of observational data. 

The results presented in this section are in very good agreement with the previous calculations in Kattawar \& Adams (\cite{kattawar71}), Stam et al. (\cite{stam06}) and  Schmid et al. (\cite{schmid06a}).

\subsection{Phase curves}
\label{sec:phcurve}

\begin{figure*}
         \resizebox{\hsize}{!}{\includegraphics{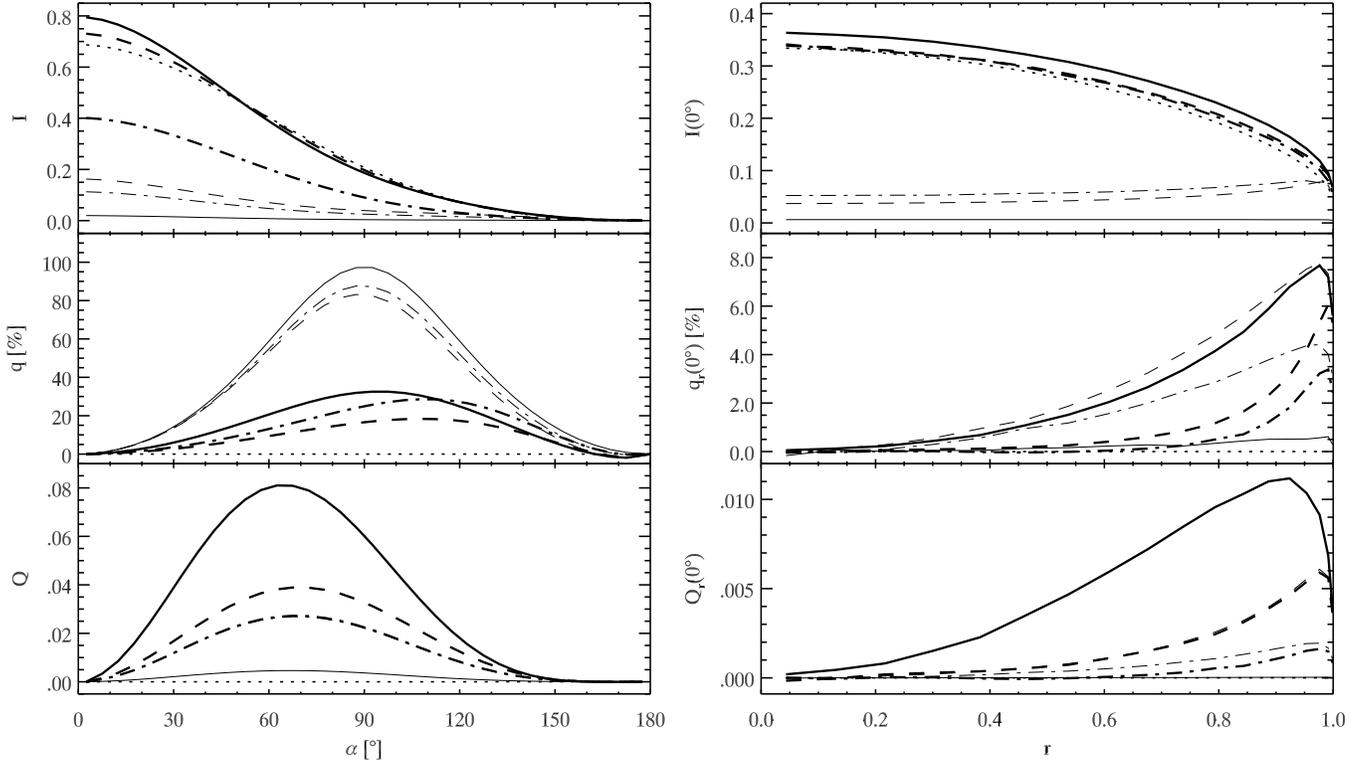}}
          \caption{Left: Phase dependence of the intensity $I$, fractional
            polarization $q$ and polarized intensity $Q$ for Rayleigh
            scattering atmospheres. Right: Radial dependence of the intensity $I$, radial
            polarization $q_r$ and radial polarized intensity $Q_r$ at
            opposition. Line styles denote: Semi-infinite case $\tau_{\mathrm
            sc}=\infty$ (solid) for single scattering albedos $\omega=1$ 
            (thick), 0.1 (thin) and finite atmosphere ($\tau_{\mathrm{sc}}=0.3$) with $\omega=$1
                        (dashed) and
            0.6 (dash-dot) for surface albedos $A_\mathrm{S}=1$ (thick) and 0 (thin). Also shown is
                         the intensity curve for conservative semi-infinite isotropic scattering
            (dotted).}
          \label{fig:plotphaseradial} 
\end{figure*}

For the investigation of extrasolar planets, the phase dependence of the disk-integrated
polarization is of interest. We discuss the phase curve for 
selected model cases (Fig.~\ref{fig:plotphaseradial}, left): a semi-infinite and a finite scattering
layer with different absorption properties of the scattering and surface layers.  

The semi-infinite, conservatively scattering layer is a good
reference case for an illuminated sphere and is often used for scattering atmospheres. All irradiated light is reflected after one or
several scatterings and the spherical albedo is equal to 1. An intensity phase curve for isotropic scattering is given in van de Hulst~(\cite{vandehulst80}), and Bhatia \&
Abhyankar (\cite{bhatia82}) published a polarization curve for Rayleigh scattering in
graphical form, but no tabulated values could be found in the literature. 

In our Monte Carlo simulation we treat 
the semi-infinite atmosphere as $\tau_{\mathrm{sc}}=30$ and $A_\mathrm{S}=1$, which yields essentially the same 
results as an infinite layer but avoids infinite scattering of some
photons. Our results of this case are tabulated in Table \ref{tab:raytau30}.

\begin{table}
\caption{Reflectivity $I(\alpha)$, polarization fraction $q(\alpha)$ and polarized intensity $Q(\alpha)$ phase curves for a very deep 
   ($\tau=30$) conservative $(\omega=1)$ Rayleigh scattering atmosphere above a perfectly
  reflecting Lambert surface (surface albedo $A_\mathrm{S}=1$). This model approximates well a conservative, semi-infinite Rayleigh scattering atmosphere. Additionally the fit parameter $a(\alpha)$ for the parametrization of the polarized intensity $Q(\alpha)$ (Eq. \ref{eq:Qfit}) is given for relevant phase angles.}
\label{tab:raytau30}
\centering
\begin{tabular}{rcrcc}
\hline\hline
$\alpha$ [$^\circ$] & $I(\alpha)$ & $q(\alpha)$ [\%]& $Q(\alpha)$ & $a(\alpha)$ \cr
\hline
2.5     &       0.795   &       0.0     &       0.0000  &               \cr
7.5     &       0.785   &       0.4     &       0.0031  &               \cr
12.5    &       0.766   &       1.1     &       0.0084  &               \cr
17.5    &       0.740   &       2.1     &       0.0155  &               \cr
22.5    &       0.708   &       3.4     &       0.0241  &       1.85    \cr
27.5    &       0.671   &       5.1     &       0.0342  &       1.86    \cr
32.5    &       0.630   &       6.9     &       0.0435  &       1.87    \cr
37.5    &       0.587   &       9.1     &       0.0534  &       1.89    \cr
42.5    &       0.542   &       11.4    &       0.0618  &       1.91    \cr
47.5    &       0.497   &       13.9    &       0.0691  &       1.94    \cr
52.5    &       0.453   &       16.6    &       0.0752  &       1.98    \cr
57.5    &       0.410   &       19.3    &       0.0791  &       2.03    \cr
62.5    &       0.368   &       22.0    &       0.0810  &       2.08    \cr
67.5    &       0.329   &       24.6    &       0.0809  &       2.14    \cr
72.5    &       0.292   &       27.0    &       0.0788  &       2.21    \cr
77.5    &       0.259   &       29.1    &       0.0754  &       2.29    \cr
82.5    &       0.228   &       30.7    &       0.0700  &       2.39    \cr
87.5    &       0.199   &       31.9    &       0.0635  &       2.49    \cr
92.5    &       0.174   &       32.5    &       0.0566  &       2.62    \cr
97.5    &       0.150   &       32.5    &       0.0488  &       2.77    \cr
102.5   &       0.130   &       31.8    &       0.0413  &       2.95    \cr
107.5   &       0.111   &       30.5    &       0.0339  &       3.16    \cr
112.5   &       0.094   &       28.6    &       0.0269  &       3.42    \cr
117.5   &       0.079   &       26.2    &       0.0207  &       3.76    \cr
122.5   &       0.066   &       23.4    &       0.0154  &               \cr
127.5   &       0.054   &       20.3    &       0.0110  &               \cr
132.5   &       0.043   &       17.0    &       0.0073  &               \cr
137.5   &       0.033   &       13.7    &       0.0045  &               \cr
142.5   &       0.025   &       10.4    &       0.0026  &               \cr
147.5   &       0.018   &       7.3     &       0.0013  &               \cr
152.5   &       0.013   &       4.4     &       0.0006  &               \cr
157.5   &       0.008   &       2.0     &       0.0002  &               \cr
162.5   &       0.005   &       0.0     &       0.0000  &               \cr
167.5   &       0.002   &       -1.4    &       0.0000  &               \cr
172.5   &       0.001   &       -1.9    &       0.0000  &               \cr
177.5   &       0.000   &               &               &               \cr     
\hline                                                           
\end{tabular}
\begin{list}{}{}
\item[] The statistical error of the Monte Carlo calculation for $I(\alpha)$ is smaller than 0.001 for all $\alpha$. The uncertainty of the polarization fraction is less than 0.1 \% for phase angles between 5 and 165 degrees. Extrapolating the intensity $I$ towards $\alpha=0^\circ$ with a quadratic least-squares fit to the first four points ($\alpha = 2.5^\circ,\ldots, 17.5^\circ$) yields a value $I(0^\circ) = 0.7970$. This agrees with the exact solution $I(0^\circ)=0.7975$ from Prather (\cite{prather74}) to the third digit. 
\end{list}           
\end{table}

\paragraph{Intensity:}
The intensity phase curves $I(\alpha)$ have their maximum at $\alpha=0^\circ$, when the whole illuminated hemisphere is visible, and they decrease steadily to zero at $\alpha=180^\circ$, where only the 
dark side of the planet is seen. The intensity $I(0^\circ)$ is equivalent 
to the geometric albedo. It is 0.7975 for the semi-infinite Rayleigh scattering atmosphere (Prather \cite{prather74}), higher than for the semi-infinite isotropic scattering model ($I_{\mathrm{iso}}(0^\circ)=0.690$, van de Hulst \cite{vandehulst80}) or a white Lambertian sphere 
($I_{\mathrm{Lam}}(0^\circ)=2/3$), because the Rayleigh scattering phase matrix favors
forward and backward scattering. On the other hand the Rayleigh scattering intensity curve
is lower for the range $\alpha\approx 52^\circ - 120^\circ$. 

Of course, the reflected intensity decreases with absorption 
(with lower single scattering albedo $\omega$) in the atmosphere and with 
the albedo $A_\mathrm{S}$ of the underlying surface layer. The effect of 
absorption in the scattering atmosphere is important for thick layers, 
while the albedo of the underlying surface is important if the optical
depths of the scattering region above is small. A quantitative 
description of these dependences is given in Dlugach \& Yanovitskij (\cite{dlugach74}) and Sromovsky (\cite{sromovsky05b}).

\paragraph{Polarization fraction.}
The disk-integrated polarization fraction $q(\alpha)$ is always zero for phase angles 
$\alpha=0^\circ$ and $\alpha=180^\circ$ because of the imposed rotational
symmetry. The polarization maximum is near the right-angle scattering configuration $\alpha\approx 90^\circ$. 
 
The polarization for the semi-infinite, conservative Rayleigh scattering layer
reaches a maximum of $q=32.6$\% for $\alpha=95^\circ$.
For reduced scattering albedo, e.g. due to absorption in a molecular band, $q(\alpha)$ increases (see e.g. van de Hulst \cite{vandehulst80}). This happens because absorption strongly 
reduces the fraction of multiply scattered photons in the reflected light which have randomized
polarization directions. If the absorption is
very strong then the reflected light consists essentially only  
of photons that made one single Rayleigh scattering. The polarization phase curve
then approaches the Rayleigh scattering
polarization phase function $p(\alpha)= (1-\cos^2\alpha)/(1+\cos^2\alpha)$ with a polarization close to 100~\% at $\alpha=90^\circ$. 

For finite scattering atmospheres the polarization fraction $q$
also depends on the albedo of the surface layer. In the models
discussed in this section the polarization is only produced in the
Rayleigh scattering layer, while reflection from the surface layer is
unpolarized. Therefore the resulting polarization  
is low for a high surface albedo and high for a low surface albedo (see e.g. Kattawar \& Adams~\cite{kattawar71}, Stam \cite{stam08}). 
This reflects the relative contribution of the polarized light from
the scattering layer with respect to  
the unpolarized light reflected from the surface underneath. 

The peak of the polarization curve is shifted towards large phase
angles ($\alpha\approx 110^\circ$) for models with thin scattering layers and 
high surface albedos, as previously described by Kattawar \& Adams
(\cite{kattawar71}). At large
phase angles ($\alpha>90^\circ$), when only a planet crescent is visible, the fraction of
scattered photons hitting the planet initially under grazing
incidence is relatively high. For a thin scattering layer grazing
incidence helps to enhance the probability for a polarizing Rayleigh scattering. 
For this reason the polarized
light from the Rayleigh scattering atmosphere is less diluted by
unpolarized light reflected from the surface at large phase angles and
the fractional polarization is higher. 

\paragraph{Polarized intensity.} The polarized intensity $Q(\alpha)$, which is
the product of polarization $q$ and intensity $I$, is zero at $\alpha=0^\circ$
and $180^\circ$, while the maximum of the phase curve $Q(\alpha)$ is
near $\alpha\approx 65^\circ$, depending slightly on the model parameters.  
The maximum value for the polarized intensity, considering the
entire parameters space, is 
$Q_{\mathrm{max}} = 0.0812$ for the semi-infinite, conservative Rayleigh 
scattering atmosphere at $\alpha=65^\circ$.
It seems unlikely that another type of scattering process and model atmosphere can produce 
a higher polarized intensity.

The polarized intensity decreases with increasing absorption, because the drop in
intensity is stronger than the increase in fractional polarization. The 
polarization flux is a rough measure for the number of reflected photons undergoing one single Rayleigh scattering. Second and higher order scatterings also add to the polarized intensity, but only at a 
much lower level. Adding absorption can only reduce the number of such scatterings and therefore diminishes the polarized intensity. 

A very important property of the polarization flux $Q$ is that it
does not depend on the albedo of the surface layer $A_\mathrm{S}$ (assumed to produce no
polarization) below the scattering region.

\subsection{Radial dependence for resolved planetary disks at opposition}
\label{sec:opposition}

For the interpretation of the limb polarization of solar system objects close to opposition, we discuss the radial or center-to-limb dependence of the 
intensity $I(r)$, the radial polarization $q_r(r)$ and the radial
polarized intensity $Q_r(r)$ (Fig.~\ref{fig:plotphaseradial}, right) for the same
model parameters as for the phase curves in
Sect.~\ref{sec:phcurve}.

\paragraph{Intensity:}
The radial intensity curve $I(r)$ shows a pronounced limb darkening in the semi-infinite
conservative case. For a strongly absorbing atmosphere, e.g. within an
absorption band, the $I(r)$-curve becomes essentially flat. Thus for an absorbing (and homogeneous)
semi-infinite atmosphere limb brightening cannot be produced. For comparison the center-to-limb intensity curve $I(r)$ for isotropic scattering and for a Lambert sphere ($I(r)= 1/\pi (1-r^2)^{1/2}$)
are also shown.

For finite scattering atmospheres with an optically thin layer the
center-to-limb intensity curve can show a limb brightening
effect. Limb brightening occurs for a highly reflective scattering
layer (high $\omega$) located above a dark surface (low $A_\mathrm{S}$), e.g. a thin aerosol layer or a methane-poor layer above the methane-rich absorbing layer (e.g. Price~\cite{price78}). Limb brightening is observed in solar system planets in deep absorption band (e.g. Karkoschka~\cite{karkoschka01}; Sromovsky \& Fry \cite{sromovsky07}). Limb brightening is investigated in more detail in Sect.~\ref{sec:limb}.

\paragraph{Radial polarization fraction:}
The radial polarization fraction $q_r(r)$ is always zero in the disk center
because of the symmetry of the scattering situation. For all cases the
polarization increases steadily towards the limb and reaches a maximum
value close to the limb between $r=0.95$ and $1.0$. The polarization
$q_r(r)$ is always positive, which means a radial polarization
direction or limb polarization perpendicular to the limb. 

It is important to note that the limb polarization decreases with
decreasing single scattering albedo $\omega$ (more absorption) in
contrast to the situation at large phase angles. 
This indicates that the photons producing the limb 
polarization are more strongly reduced by
absorption than the reflected ``unpolarized'' photons.   

The explanation is that singly scattered (i.e. backscattered) photons
do not contribute to the limb
polarization, while reflected photons scattered twice or a few times are responsible  
for the largest part of the limb polarization. Absorption implies
that a larger fraction of escaping photons are singly-scattered and therefore
unpolarized at opposition. Note however that for the semi-infinite atmosphere,
the maximum radial polarization is not reached in the conservative
case. A slightly lower scattering albedo ($\omega \approx 0.95$) 
mostly reduces the amount of highest order scatterings 
and thus the polarization fraction is somewhat enhanced when 
compared to the conservative case
(see Schmid et al.~\cite{schmid06a}, 
and Fig.~\ref{fig:fp0_ray} in Sect.~\ref{sec:param}). 

The fractional limb polarization $q_r(r)$ for finite scattering layers depends
strongly on the albedo of the underlying surface $A_\mathrm{S}$: $q_r(r)$ is high for low $A_\mathrm{S}$ and low for high $A_\mathrm{S}$ like for large phase angles. 
A low surface albedo decreases the photons with multiple scatterings in the plane perpendicular to the limb, which are polarized parallel to the limb, thus enhancing the polarization in perpendicular direction. Therefore the limb polarization of a bright layer over a dark one can be even higher than for a semi-infinite atmosphere. This is discussed in more detail in Sect.~\ref{sec:param}.

\paragraph{Radial polarized intensity:} The radial polarized
intensity $Q_r(r) = q_r(r) \cdot I(r)$ increases with $r$ from
zero in the disk center to a maximum at $r>0.9$ and then drops
at the very limb. For semi-infinite atmospheres, $Q$ just 
decreases at all radii with decreasing single scattering albedo $\omega$.

For finite atmospheres, the limb polarization flux $Q_r(r)$ 
depends only slightly on the surface albedo. Decreasing $A_\mathrm{S}$ from 1 to 0 can increase $Q_r(r)$ at most $\sim0.002$ for some models, while for most models $Q_r(r)$ is virtually constant. This is similar to the case for large phase angles.

\subsubsection{Limb darkening and limb brightening vs. limb polarization} \label{sec:limb}

\begin{figure}
         \resizebox{\hsize}{!}{\includegraphics{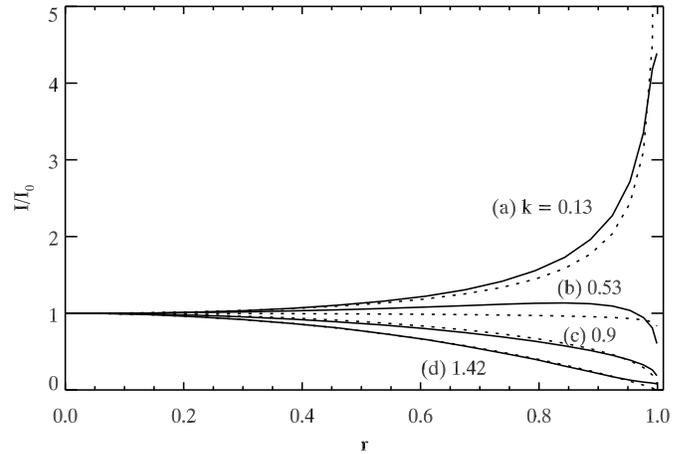}}
          \caption{Radial intensity curves for different model parameters normalized to the central disk intensity, showing examples of limb darkening and limb brightening. The solid line is the calculated model and the dotted line the best fit with the Minnaert law. (a), (b) and (c) are for conservatively ($\omega=1$) scattering layers above black surfaces ($A_\mathrm{S}=0$) with scattering layer thickness $\tau_{\mathrm{sc}} = 0.1$ (a), 1 (b) and $\infty$ (c). (d) is a thin ($\tau_{\mathrm{sc}} = 0.2$), highly absorbing $\omega=0.2$ scattering layer above a white surface ($A_\mathrm{S} = 0$).} 
          \label{fig:limblaw} 
\end{figure}

For a surface with a low albedo $A_\mathrm{S}$ below a thin scattering layer the limb can be brighter than the disk
center, an effect that is generally called limb brightening. In principle this effect should be 
called ``a central disk darkening'', because the low surface albedo $A_\mathrm{S}$ does not brighten the limb.
It only absorbs more light in the center of the disk, where a higher fraction of
photons reach the absorbing surface because of their perpendicular
incidence when compared to the situation of grazing incidence at the limb.
Despite this fact we will retain the term ``limb brightening'' and consider the
limb brightness on a relative scale compared to the brightness of the
disk center. 

The limb darkening and limb brightening effect can be parametrized to a first order
using the Minnaert law $I(r) = I_{\mathrm{r=0}} (1-r^2)^{k-1/2}$. The Minnaert
parameter k determines the shape of the curve, $k=1$ corresponds to
Lambert's law, $k=0.5$ to a flat intensity distribution $I(r) = I_0$ and $k < 0.5$
to limb brightening. The k parameter was determined by fitting
Minnaert's law to all modeled intensity profiles, fixing the
intensity at the center and excluding the outermost point where the
formula diverges for $k<0.5$. With the exception of some cases
mentioned below, most profiles can be fitted adequatly.

In Fig.~\ref{fig:limblaw} different examples of limb darkening and
limb brightening are shown along with the best fit of the Minnaert
law. Intensities are normalized to the central disk intensity. 
There are two types of limb brightening curves. For very thin
atmospheres the maximal brightness is measured at the very edge of the
planet, while for a moderate optical depth the intensity raises slightly up to
a certain radius (e.g. $0.9 R_p$) and then drops very close to the
limb. This second case cannot be fitted with the one-parameter
Minnaert law and is approximated here by a relatively flat curve
$k\approx0.5$. 

The limiting case of a conservative semi-infinite atmosphere yields a Minnaert parameter of
$k\approx0.9$. For $\omega$ going towards $0$, $k$ tends to a flat intenstiy distribution $k=0.5$. For finite atmospheres there
is a strong dependence on the surface albedo $A_\mathrm{S}$. For a strongly absorbing atmosphere over a bright surface ($\omega$ low,
$A_\mathrm{S}$ high) absorption is more likely towards the limb ($k > 1$), for a bright atmosphere
over a dark surface the opposite is true ($k < 0.5$). In the latter case the central disk intensity
is very low. 

Similar to limb brightening, the limb polarization is also enhanced for a bright scattering layer
over a dark surface. However, there are fundamental differences between these two effects. Limb polarization
arises only for a polarizing process like Rayleigh scattering, while limb brightening occurs also for non-polarizing scattering processes like isotropic scattering. Additionally limb brightening is the stronger the thinner the upper bright layer, while
limb polarization requires a sufficiently thick scattering layer above the dark surface. Finally, limb polarization
can also occur for cases of limb darkening, e.g. the semi-infinite, conservative atmosphere. Therefore, the two 
effects provide complementary diagnostics of the vertical structure of the atmosphere.

This section explores the full parameter space for 
simple Rayleigh scattering atmospheres. We explore the parameter space by varying one of the three parameters $\omega$, $\tau_{\mathrm{sc}}$ and $A_\mathrm{S}$ (cf. sec. \ref{sec:atmosphere}) while fixing the other two. We study the resulting intensity $I(90^\circ)$, polarization fraction $q(90^\circ)$, and polarized intensity $Q(90^\circ)$ (Figs. \ref{fig:fp90_ray} to \ref{fig:fp90_agr}).

The shapes of the model phase curves for the intensity $I(\alpha)$,
fractional polarization $q(\alpha)$, and polarized intensity $Q(\alpha)$ 
look very similar for different 
model parameters (see Fig.~\ref{fig:plotphaseradial}). 
Therefore it is reasonable for a model parameter study to select  
the results for the phase angle $\alpha=90^\circ$, 
considering them as representative (qualitatively) for all phase angles. 
A phase angle $\alpha=90^\circ$ is ideal for extrasolar planets because all
planets will pass through this configuration twice during an orbit,
regardless of inclination.

The same type of parameter study is presented for the limb
polarization of planets at opposition ($\alpha = 0^\circ$). For this we determine
disk-integrated (averaged) quantities for the intensity
and radial polarization from the model results (Figs. \ref{fig:fp0_ray} to \ref{fig:fp0_agr}). The integrated intensity $I(0^\circ)$ is
equivalent to the geometric albedo, $\langle q_r\rangle$ is the
intensity weighted average of the fractional polarization, 
and $\langle Q_r\rangle$ the
integrated polarized intensity on the same scale as the integrated
intensity. These quantities are determined as described in
Sect.~\ref{sec:polpar}.  

Figures \ref{fig:fp90_ray} and \ref{fig:fp0_ray} show the dependence
on the single scattering albedo $\omega$. For a given scattering
optical depth $\tau_{\mathrm{sc}}$ a reduction in $\omega$ is equivalent to an
enhancement of the absorption $\kappa$ in the scattering layer. 
Strong differences in $\kappa(\lambda)$ occur in planetary atmospheres for
molecular absorptions (e.g. due to CH$_4$ or H$_2$O) inside and
outside the band while $\sigma$ is essentially equal. 

In Figs. \ref{fig:fp90_tau} and \ref{fig:fp0_tau} the Rayleigh
scattering optical depths from $\tau_{\mathrm{sc}} = 10.0$ to $0.01$ are plotted. 
This illustrates quite well the possible
spectral dependence from short to long wavelengths (left to right) 
for a Rayleigh scattering atmosphere. Since the Rayleigh
scattering cross sections is proportional to $1/\lambda^4$, it is
possible that a planet has $\tau_{\mathrm{sc}} = 4$ at 400 nm and $\tau_{\mathrm{sc}} = 1/4$
at 800 nm.

\subsection{Parameter study for quadrature phase $\alpha=90^\circ$ and opposition $\alpha=0^\circ$}
\label{sec:param}

\begin{figure}[p!]
         \resizebox{\hsize}{!}{\includegraphics{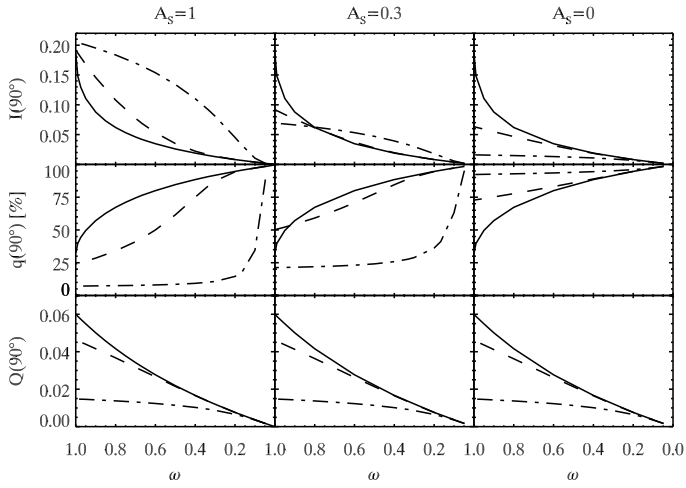}}
          \caption{Intensity, polarization and polarized intensity at quadrature as function of single scattering albedo $\omega$ for optical depths $\tau_{\mathrm{sc}}=\infty$ (solid), 0.6 (dashed), 0.1 (dash-dot) and surface albedos $A_\mathrm{S} = 1$ (left), 0.3 (middle), 0 (right).} 
          \label{fig:fp90_ray} 
\end{figure}

\begin{figure}[p!]
         \resizebox{\hsize}{!}{\includegraphics{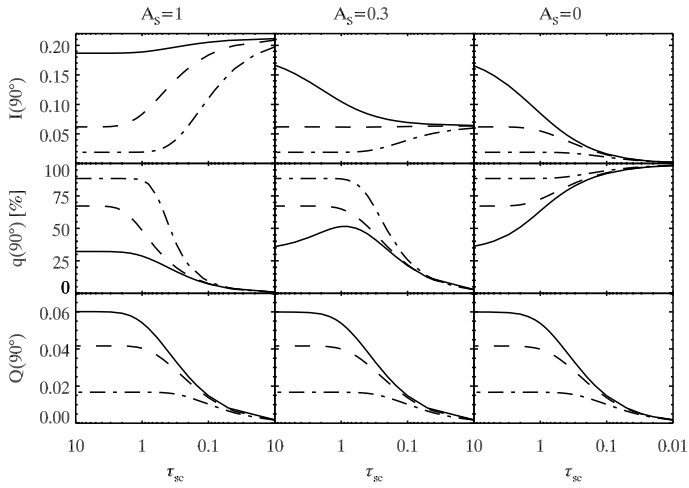}}
          \caption{Intensity, polarization and polarized intensity at quadrature as function of optical depth $\tau_{\mathrm{sc}}$ for single scattering albedos $\omega=1$ (solid), 0.8 (dashed), 0.4 (dash-dot) and surface albedos $A_\mathrm{S} = 1$ (left), 0.3 (middle), 0 (right).} 
          \label{fig:fp90_tau} 
\end{figure}

\begin{figure}[p!]
         \resizebox{\hsize}{!}{\includegraphics{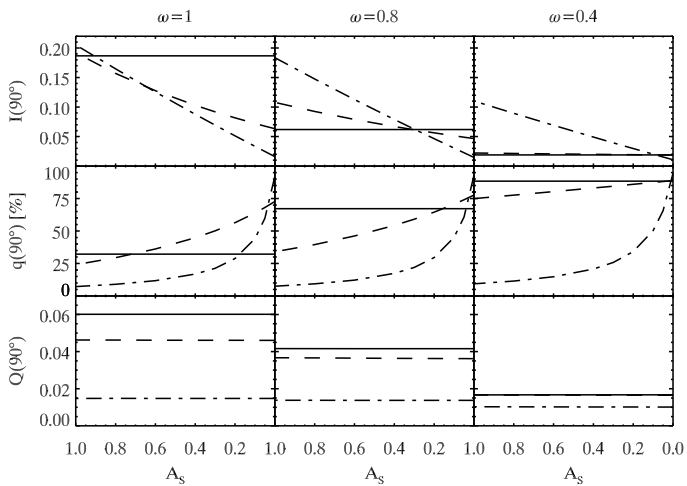}}
          \caption{Intensity, polarization and polarized intensity at quadrature as function of surface albedo $A_\mathrm{S}$ for optical depths $\tau_{\mathrm{sc}}=\infty$ (solid), 0.6 (dashed), 0.1 (dash-dot) and single scattering albedos $\omega=\omega_{\mathrm{Ray}}=1$ (left), 0.8 (middle), 0.4 (right).} 
          \label{fig:fp90_agr} 
\end{figure}

\begin{figure}[p!]
         \resizebox{\hsize}{!}{\includegraphics{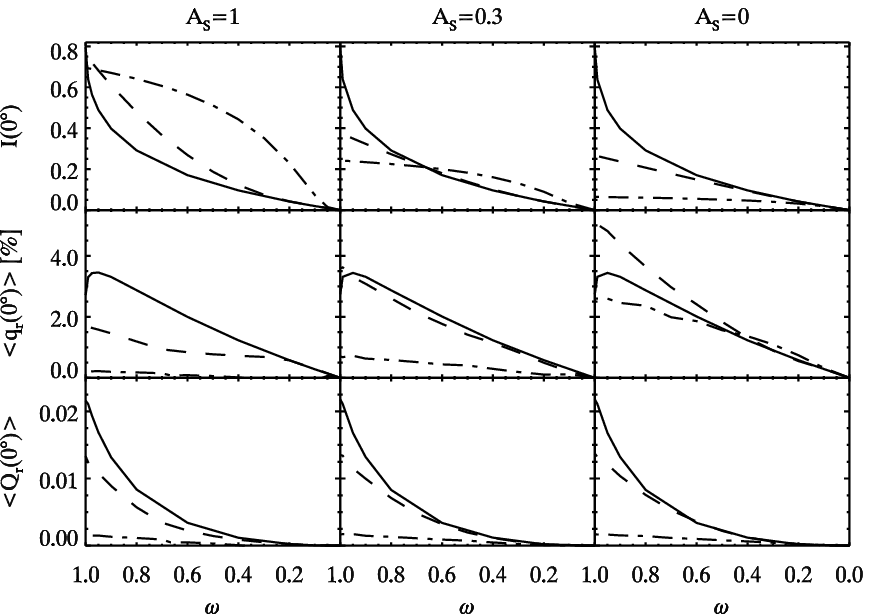}}
          \caption{Geometric albedo, disk-integrated radial polarization and polarized intensity at opposition as function of single scattering albedo $\omega$ for optical depths $\tau_{\mathrm{sc}}=\infty$ (solid), 0.6 (dashed), 0.1 (dash-dot) and surface albedos $A_\mathrm{S} = 1$ (left), 0.3 (middle), 0 (right).} 
          \label{fig:fp0_ray} 
\end{figure}

\begin{figure}[p!]
         \resizebox{\hsize}{!}{\includegraphics{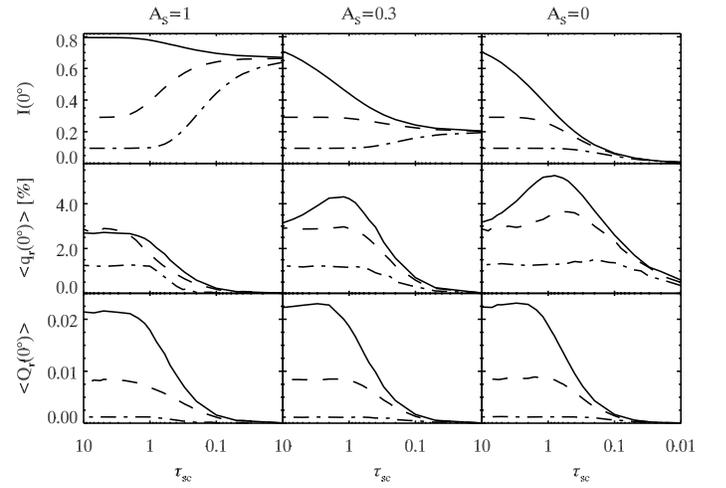}}
          \caption{Geometric albedo, disk-integrated radial polarization and polarized intensity at opposition as function of optical depth $\tau_{\mathrm{sc}}$ for single scattering albedos $\omega=1$ (solid), 0.8 (dashed), 0.4 (dash-dot) and surface albedos $A_\mathrm{S} = 1$ (left), 0.3 (middle), 0 (right).} 
          \label{fig:fp0_tau} 
\end{figure}

\begin{figure}[p!]
         \resizebox{\hsize}{!}{\includegraphics{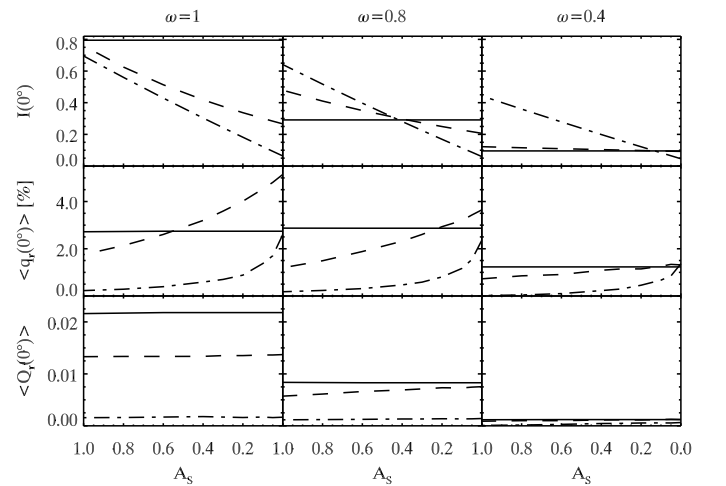}}
          \caption{Geometric albedo, disk-integrated radial polarization and polarized intensity at opposition as function of surface albedo $A_\mathrm{S}$ for optical depths $\tau_{\mathrm{sc}}=\infty$ (solid), 1 (dotted), 0.6 (dashed), 0.1 (dash-dot) and single scattering albedos $\omega=\omega_{\mathrm{Ray}}=1$ (left), 0.8 (middle), 0.4 (right).} 
          \label{fig:fp0_agr} 
\end{figure}

\parskip 10pt

The effect of the albedo $A_\mathrm{S}$ of the surface below the Rayleigh
scattering layer is shown in Figs.~\ref{fig:fp90_agr} and
\ref{fig:fp0_agr}.  

General results from the Figures \ref{fig:fp90_ray} to
\ref{fig:fp0_agr} are:
\begin{itemize}
\item{} lowering the Rayleigh scattering albedo $\omega$
  always results in a lower intensity $I$, and lower
  polarized intensity $Q$ or $Q_r$,
\item{} lowering the Rayleigh scattering albedo $\omega$ results in a 
        higher polarization $q$ at large phase angles. Contrary to this the
  fractional limb polarization $q_r$ is reduced for lower $\omega$,  
\item{} lowering the Rayleigh scattering optical depth $\tau_{\mathrm{sc}}$
  produces a strong reduction in
  the polarized intensity $Q$ or $Q_r$ in the optically thin case $\tau \lesssim 2$ and causes
  essentially no change in $Q$ or $Q_r$ in the optical thick case $\tau \gtrsim 2$, 
\item{} lowering the surface albedo $A_\mathrm{S}$ lowers the intensity $I$ and
  enhances the fractional polarization $q$ or $q_r$,
\item{} changing the surface albedo $A_\mathrm{S}$ does not change the
  polarized flux $Q$ and hardly $Q_r$.
\end{itemize}

The most important difference between the limb polarization $\langle q_r \rangle$ and the
disk-averaged polarization $q(90^\circ)$ is their opposite
dependence on the Rayleigh scattering albedo $\omega$ (see e.g. the
middle panels of Figs.~\ref{fig:fp90_agr} and \ref{fig:fp0_agr}).
This occurs because the limb polarization at
opposition is mainly caused by photons undergoing two to about six scatterings
rather than just one.

Another difference is the influence of $\tau_{\mathrm{sc}}$ on the
fractional polarization: $q$ drops with $\tau_{\mathrm{sc}}$ for bright non-polarizing
surfaces and increases for dark surfaces. It is more complicated at opposition: the limb polarization is highest if the dark ground eliminates photons that would otherwise scatter twice perpendicular to the limb, but the atmosphere is still thick
enough to produce many photons that escape having scattered twice parallel to the limb. 
The maximum possible limb polarization $\langle q_r \rangle = 5.25$\% is reached for
$\tau_r=0.8$, $A_\mathrm{S}=0$ and $\omega=1$.

From the variation of $\tau_{\mathrm{sc}}$ shown in Fig.~\ref{fig:fp90_tau} it can
be seen that the polarized intensity $Q(90^\circ)$ saturates above $\tau = \tau_{\mathrm{sc}}\cdot \omega \gtrsim 2$. 
Therefore $Q(90^\circ)$ cannot probe deep atmospheric layers. 
For the intensity and fractional polarization, an absorbing ground under a conservatively
scattering layer can be noticed even at $\tau\gtrsim10$. 

The polarized intensity $Q(\alpha)$ consists mostly of photons undergoing just one single Rayleigh scattering. Therefore, $Q$ is not
changed by processes which happen deep in the atmosphere or by diffuse scattering on the surface. $Q$ is only reduced if the number of single Rayleigh scatterings are reduced, e.g. because there is only a thin Rayleigh scattering layer or photons are efficiently absorbed high in the
atmosphere.
  
We can approximate the polarized intensity $Q$ 
by the following parametrization:

\begin{equation}\label{eq:Qfit}
Q (\alpha,\tau_{\mathrm{sc}},\omega) = Q(\alpha,\infty,1)\cdot(1-\mathrm{e}^{-a(\alpha)\tau_{\mathrm{sc}}})\cdot \omega^{b(\alpha,\tau_{\mathrm{sc}})},
\end{equation}

\noindent where $a(\alpha)$ and $b(\alpha,\tau_{\mathrm{sc}})$ are fit parameters.  Table~\ref{tab:pffitparam} shows the best fit parameter $b(\alpha,\tau_{\mathrm{sc}})$, while $Q(\alpha,\infty,1)$ and $a(\alpha)$ are listed in Table~\ref{tab:raytau30}.
 
\begin{table}
\caption{Best fit parameter $b(\alpha,\tau_{\mathrm{sc}})$ for the parametrization of the polarized intensity Q (Eq. \ref{eq:Qfit}).}
\label{tab:pffitparam}
\centering
\begin{tabular}{rcccc}
\hline\hline
$\tau_{\mathrm{sc}}$ & $b(30^\circ,\tau_{\mathrm{sc}})$ & $b(60^\circ,\tau_{\mathrm{sc}})$ & $b(90^\circ,\tau_{\mathrm{sc}})$ & $b(120^\circ,\tau_{\mathrm{sc}})$ \cr
\hline
0.1             &       0.38    &       0.39    &       0.39    &       0.44    \cr
0.2             &       0.63    &       0.65    &       0.66    &       0.75    \cr
0.3             &       0.77    &       0.80    &       0.82    &       0.90    \cr
0.5             &       0.99    &       1.02    &       1.04    &       1.10    \cr
0.8             &       1.23    &       1.22    &       1.26    &       1.23    \cr
1.0             &       1.32    &       1.33    &       1.32    &       1.25    \cr
2.0             &       1.52    &       1.48    &       1.42    &       1.27    \cr
10.0    &       1.59    &       1.59    &       1.45    &       1.28    \cr
\hline                                                                   
\end{tabular}                                                                                                                    
% \noindent                                                                                                                      
\end{table}

For optically thick Rayleigh scattering atmospheres the polarized intensity
$Q$ depends only on the single scattering albedo $\omega$, and the parametrization reduces to:
\begin{equation} \label{eq:pf_alpha}
Q(\alpha,\omega,\tau_{\mathrm{sc}}\gtrsim2) \approx Q(\alpha,\infty,1)\cdot\omega^{b(\alpha,\tau_{\mathrm{sc}}\gtrsim2)}\,.
\end{equation}
At quadrature this is
\begin{equation}
Q(90^\circ,\omega,\tau_{\mathrm{sc}}\gtrsim2) \approx 0.060 \cdot \omega^{1.45}\, (\pm 0.002)\,.
\end{equation}

For the limb polarization flux $\langle Q_r(0^\circ) \rangle$ the dependence on $\omega$ is much steeper because both $I$ and $q_r$
drop with decreasing $\omega$, as can be seen from the bottom
panels of Fig.~\ref{fig:fp0_ray}. For thick
Rayleigh scattering layers, $\tau\gtrsim2$, the $\omega$-dependence of
the limb polarization flux is 
\begin{equation}
Q_r(\tau_{\mathrm{sc}}\gtrsim2) \approx 0.022 \cdot \omega^{4.23}\, (\pm 0.001)\,.
\end{equation}

\section{Models beyond a Rayleigh scattering layer with a Lambert
  surface}
  \label{sec:beyond}
The three parameter model grid discussed in Sect.~\ref{sec:param}
provides an overview on basic
dependences of simple Rayleigh scattering models. In this 
section we describe a few results for particle scattering properties
different from pure Rayleigh scattering, or models with more than one
polarizing scattering layer.

\subsection{Atmospheres with Rayleigh and isotropic scattering}
 \label{sec:isotropic}

\begin{figure}
         \resizebox{\hsize}{!}{\includegraphics{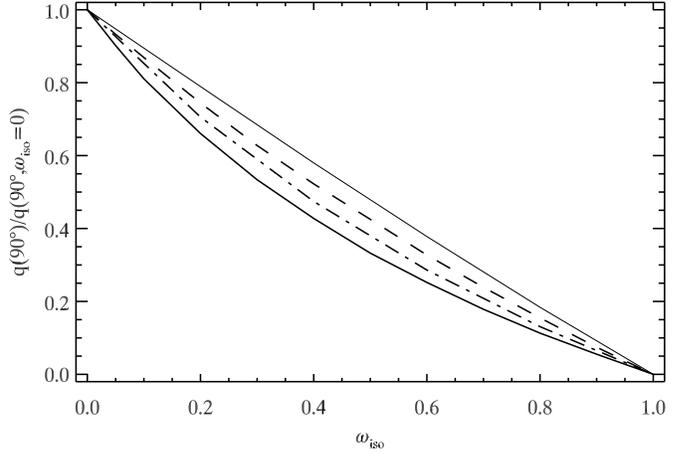}}
          \caption{Polarization of atmospheres with Rayleigh and isotropic scattering at 90$^{\circ}$ as a function of isotropic single scattering albedo $\omega_i$, normalized to the case of pure Rayleigh scattering $\omega_r = 1$ or $\omega_i=0$. Plotted models are: semi-infinite atmosphere (thick solid), $\tau_{\mathrm{sc}} = 0.3$, $A_\mathrm{S} = 1$ (dashed), $\tau_{\mathrm{sc}} = 0.3$, $A_\mathrm{S} = 0$ (dash-dotted), and $\tau_{\mathrm{sc}} = 0.05$ (thin solid). All models are without absorption, i.e. $\omega_i+\omega_r=1$.} 
          \label{fig:plotnisomodels} 
\end{figure}

Pure Rayleigh scattering is a simplification for planetary atmospheres. Already for 
Rayleigh scattering by molecular hydrogen one needs to account for a weak depolarization effect, because the diatomic molecule is non-spherical. Another depolarization effect for scattered radiation occurs in dense gas because collisions with other particles take place frequently during the scattering process. 
In addition aerosols and dust particles can also be efficient scatterers in planetary atmospheres and the net scattering phase matrix differs from Rayleigh scattering and should be evaluated, e.g. by using the more general Mie theory.

A simple way for taking such effects into account in a first approximation is to use a linear combination of
the Rayleigh scattering and isotropic scattering phase matrices
\begin{equation}
\mathbf{S} = w \cdot \mathbf{R} + (1-w)\cdot \mathbf{I}\,,    
\end{equation}
where $w=\sigma_{\mathrm{Ray}}/\sigma$ and $1-w=\sigma_{\mathrm{iso}}/\sigma$ are the relative contributions 
of the Rayleigh scattering and isotropic scattering to the total scattering cross section 
$\sigma=\sigma_{\mathrm{Ray}} + \sigma_{\mathrm{iso}}$. Note that the single
scattering albedo $\omega$ and the scattering optical depth $\tau_{\mathrm{sc}}$ now
include both the Rayleigh and the isotropic scattering cross
section (cf. Sect.~\ref{sec:atmosphere}). 

Isotropic scattering is non-polarizing. If the scattering in the
atmosphere is composed of both isotropic and Rayleigh scattering, then
the fractional polarization and the polarized intensity are reduced by isotropic scattering, while
the intensity is comparable (cf. Fig.~\ref{fig:plotphaseradial}).

Figure $\ref{fig:plotnisomodels}$ shows the fractional
polarization $p(90^\circ)$ at quadrature as a function of 
 $\sigma_{\mathrm{iso}}/\sigma$ for a few representative cases. In the single scattering limit the 
 decrease is linear, the strongest deviation from a linear law is found for the semi-infinite
atmosphere because of the large amount of multiple scatterings. A similar behavior is found for other phase
angles, as well as for the radial limb polarization at opposition.

\subsection{Forward-scattering phase functions}
\label{sec:haze}

The high polarization of Jupiter's poles and the disk-integrated polarization of Titan (e.g. Tomasko \& Smith \cite{tomasko82}; Smith \& Tomasko \cite{smith84}) has been explained by the presence of a thick layer of polarizing haze particles. The derived single scattering properties indicate strong forward scattering and Rayleigh-like linear polarization with maximal polarization close to 100\% at about 90$^\circ$ scattering angle. Particles that satisfy this behavior are thought to be aggregates that are non-spherical and with a projected area smaller than optical wavelengths (e.g. West \cite{west91}). We investigate the polarization properties of a planet with such a haze layer. The particle scattering properties are implemented as described in Braak et al. (\cite{braak02}) using a simple parametrized scattering matrix of the form

\begin{equation}
 \mathbf{F(\vartheta)} = \left(
 \begin{array}{c c c c}
 F_{11}(\vartheta) & F_{12}(\vartheta) & 0 & 0\\
 F_{12}(\vartheta) & F_{11}(\vartheta) & 0 & 0\\
 0 & 0 & F_{33}(\vartheta) & 0 \\
 0 & 0 & 0 & F_{44}(\vartheta)
 \end{array} \right)\,,
\end{equation}

\noindent where $\vartheta$ is the scattering angle and 

\begin{equation}\label{eq:hg}
F_{11}(\vartheta) = P_{HG}(g,\vartheta) = {1 - g^2 \over (1+g^2-2g\cos\vartheta)^{(3/2)}}\,,
\end{equation}
\begin{equation}\label{eq:f12_haze}
{F_{12}(\vartheta) \over F_{11}(\vartheta)} = p_m{\cos^2\vartheta-1 \over \cos^2\vartheta+1}\,,
\end{equation}
\begin{equation}\label{eq:f33_haze}
{F_{33}(\vartheta) \over F_{11}(\vartheta)} = {2\cos\vartheta \over \cos^2\vartheta + 1}\,,
\end{equation}
\begin{equation} \label{eq:f44_haze}
F_{44} = 0\,.
\end{equation}

\noindent $F_{11}(\vartheta)$ or $P_{HG}(\vartheta)$ is the Henyey-Greenstein phase function with the asymmetry parameter $g$ (see e.g. Van de Hulst \cite{vandehulst80}). $g=0$ corresponds to isotropic scattering, 
$g=1$ to pure forward scattering, $g < 0$ to enhanced backscattering. Since haze particles have been shown
to be strongly forward scattering, we limit our discussion to the two cases $g=0.6$ and $g=0.9$. 

\begin{figure}
         \resizebox{\hsize}{!}{\includegraphics{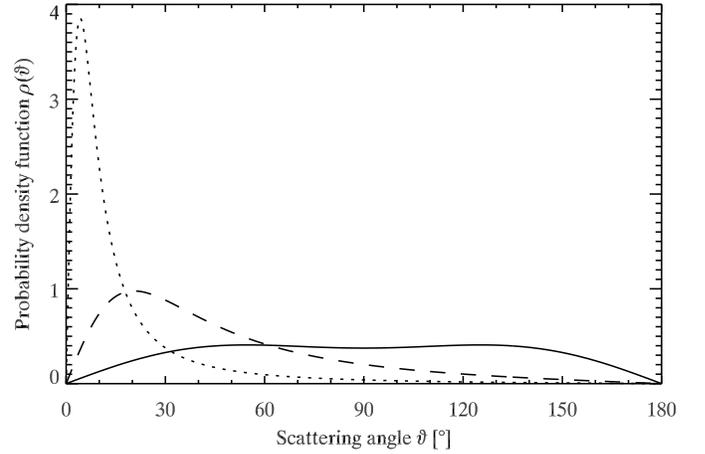}}
          \caption{Probability density function $\rho(\vartheta)$ for Rayleigh scattering (solid), Henyey-Greenstein function with asymmetry parameter $g=0.6$ (dashed) and $g=0.9$ (dotted).}
          \label{fig:phfunc} 
\end{figure}

Figure~\ref{fig:phfunc} shows the probability density function $\rho(\vartheta)$ for $P_{HG}(\vartheta)$ in comparison with Rayleigh scattering. The probability density function for the scattering angle $\vartheta$ is the phase function $F_{11}(\vartheta)$ weighted by $\sin(\vartheta)$ and normalized such that the integral over $\rho(\vartheta)$ equals 1. From this function the probability of the scattering angle within a certain interval is calculated by integrating $\rho(\vartheta)$ over this interval. One can see that for the haze models small scatttering angles (foward scattering) are greatly enhanced in comparison to Rayleigh scattering, while the probability for backscattering is much lower. 

\begin{figure*}
         \resizebox{\hsize}{!}{\includegraphics{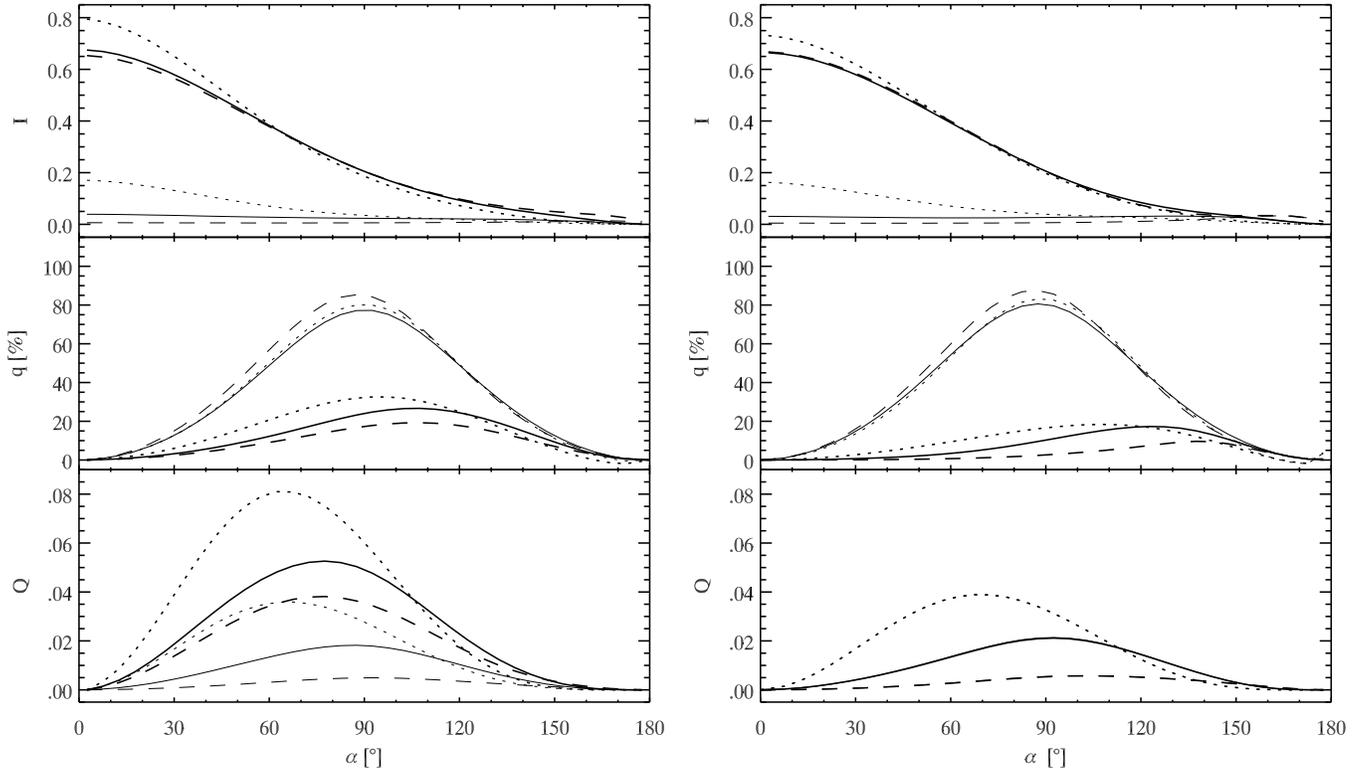}}
          \caption{Phase dependence of the intensity $I$, fractional
            polarization $q$ and polarized intensity $Q$ for a haze layer. Left: Semi-infinite case $\tau_{\mathrm
            sc}=\infty$ for single scattering albedos $\omega=1$ 
            (thick), 0.6 (thin) and $g = 0.6$ (solid), 0.9 (dashed). The dotted line is the Rayleigh scattering case for comparison. Right: Finite atmosphere $\tau_{\mathrm{sc}} = 0.3$ with $\omega=1$ for surface albedos $A_\mathrm{S}=1$ (thick), 0 (thin) and line styles as for the left plot.}
          \label{fig:phcurveshaze} 
\end{figure*}

\begin{figure*}
         \resizebox{\hsize}{!}{\includegraphics{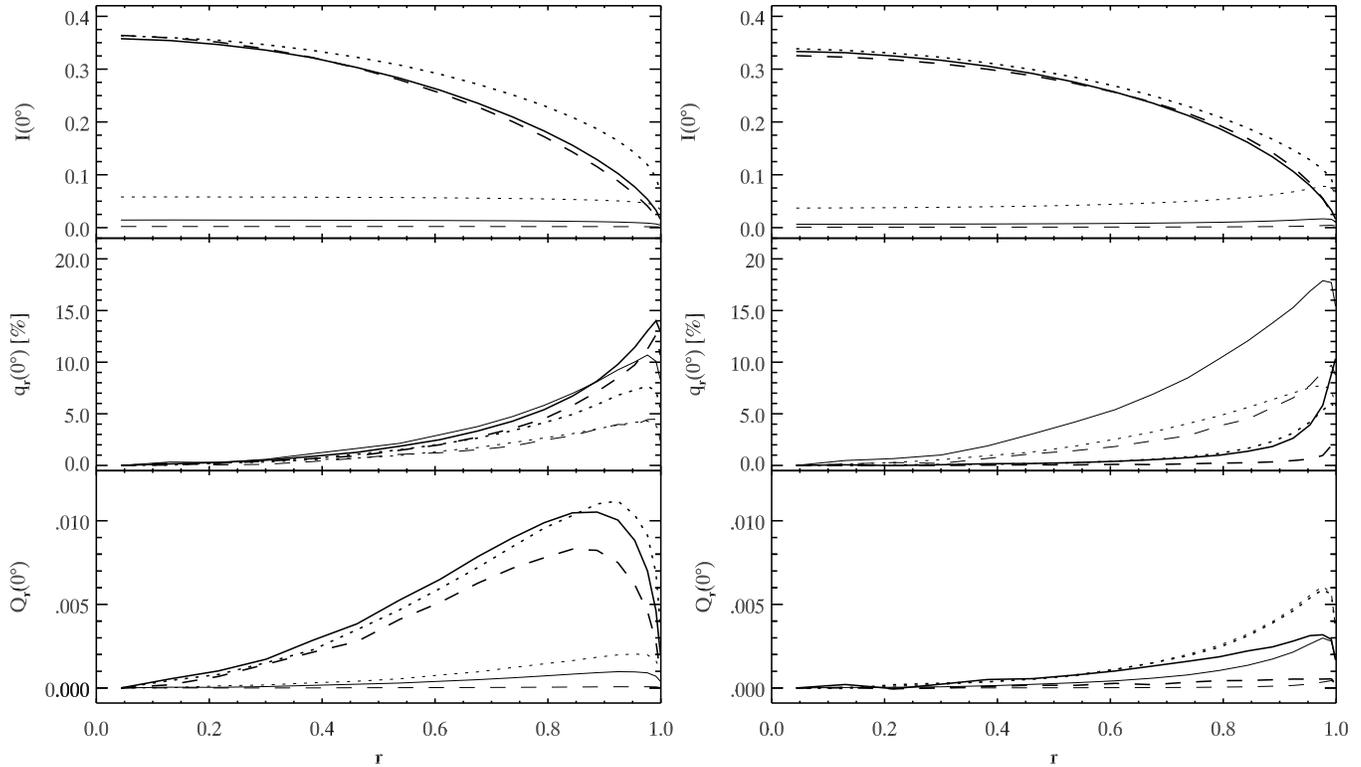}}
          \caption{Radial dependence of the intensity $I$, radial
            polarization $q_r$ and radial polarized intensity $Q_r$ at
            opposition for a haze layer. Left: Semi-infinite case $\tau_{\mathrm
            sc}=\infty$ for single scattering albedos $\omega=1$ 
            (thick), 0.6 (thin) and $g = 0.6$ (solid), 0.9 (dashed). The dotted line is the Rayleigh scattering case for comparison. Right: Finite atmosphere $\tau_{\mathrm{sc}} = 0.3$ with $\omega=1$ for surface albedos $A_\mathrm{S}=1$ (thick), 0 (thin) and line styles as for the left plot.}
          \label{fig:radialhaze} 
\end{figure*}

$F_{12}(\vartheta)/F_{11}(\vartheta)$ describes the fractional polarization of the scattered radiation as a function of the scattering angle. For scattering on haze particles it can be similar to Rayleigh scattering scaled by a factor $p_m$, the maximal single scattering polarization at $90^\circ$ scattering angle. For a first qualitative analysis we set $p_m=1$ which is an upper limit that may slightly overestimate the resulting polarization. The other matrix elements are identical to Rayleigh scattering.  

Figures~\ref{fig:phcurveshaze} and \ref{fig:radialhaze} show the phase and radial dependences for the haze models similar to the Rayleigh scattering case in Sect.~\ref{sec:phcurve} and \ref{sec:opposition}. 

\paragraph{Intensity:} The phase curves of the haze models differ from the Rayleigh scattering models mainly at small phase angles. The geometric albedo is lower for the haze models because backscattering is strongly suppressed compared to Rayleigh scattering. This is already discussed by Dlugach \& Yanovitskij (\cite{dlugach74}), who calculated albedos for semi-infinite hazy atmospheres. Our calculations result in slightly higher albedos because of the inclusion of polarization. At phase angles around 90$^\circ$ the intensities are very similar for all models for non-absorbing atmospheres. An absorber greatly reduces the albedo of a planet with enhanced forward scattering, because many photons penetrate deeply into the atmosphere after the first scattering and then have a high probability of being absorbed. The radial intensity curves mainly reflect the lower geometric albedo, while the shape of the curve is similar for all models. 

\paragraph{Polarization fraction:} The angle of maximal polarization is generally larger for the haze models than for Rayleigh scattering. In the semi-infinite conservative case it is $\approx 110^\circ$ for haze as opposed to $\approx 95^\circ$ for Rayleigh scattering. The shift to larger angles is particularly enhanced for a finite haze layer over a bright Lambert surface. However the maximal polarization decreases with increasing $g$. For strong absorption, both in or below the scattering layer, the polarization phase curve tends toward the single scattering function like in the Rayleigh case. 
The fractional limb polarization of haze layers can be much higher than for Rayleigh scattering layers, with disk-integrated values reaching $\langle q_r \rangle \approx 11$\% and peak maxima $q_r(r) \approx 20$\%. This is understandable because the singly scattered (backscattered) photons which are unpolarized are strongly reduced for foward scattering particles.  

\paragraph{Polarized intensity:} The polarized intensity $Q(\alpha)$ is significantly lower for forward scattering phase functions than for Rayleigh scattering in the phase angle range $\alpha = 30^\circ - 90^\circ$ and for the limb polarization effect at opposition. It drops strongly with increasing $g$ or increasing absorption. Like for Rayleigh scattering the polarized intensity is independent of the surface albedo $A_\mathrm{S}$. The phase curves $Q(\alpha)$ show a shift of the maximum towards larger phase angles when compared to Rayleigh scattering, in particular for models with thin scattering layers.

\subsection{Models with two polarizing layers} 

\label{sec:surface}

\begin{figure}
         \resizebox{\hsize}{!}{\includegraphics{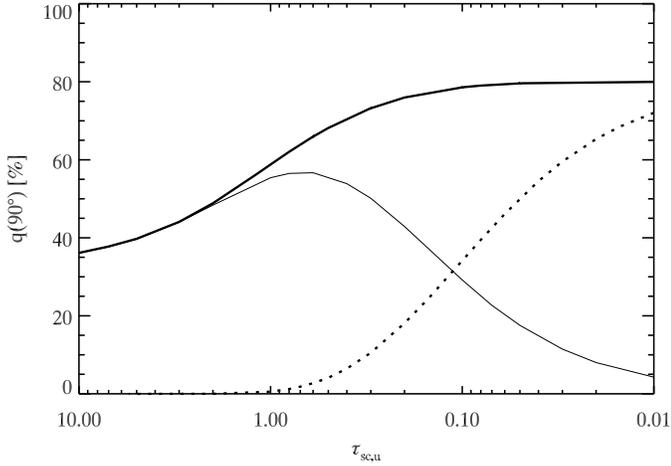}}
          \caption{Fractional polarization $q$ as function of $\tau_{\mathrm{sc}}$ of the upper layer at quadrature for models with Rayleigh scattering (solid) or isotropic (dotted) upper layer, and semi-infinite Rayleigh scattering lower layer with $\omega=0.6$ (thick) or Lambertian lower layer with  $A=0.2$ (thin), which provides the same reflectivity.} 
          \label{fig:2lp90_tau} 
\end{figure}

Up to now we have treated the region below the scattering layer 
simply as a Lambert surface with an albedo $A_\mathrm{S}$, which
produces no polarization. In this section we explore model results for
two polarizing layers with different absorption properties, where the lower layer can be a 
semi-infinite Rayleigh scattering atmosphere as described in
Sect.~\ref{sec:phcurve}. 

We focus on the question at what depth of the upper scattering layer 
$\tau_{\mathrm{sc}}$ the polarization properties of the underlying layer are washed
out by multiple scattering and are no longer recognizable in the reflected
radiation.   

Figure \ref{fig:2lp90_tau} compares the fractional polarization $q(90^\circ)$ for three cases as a function of scattering optical depth of the upper layer, $\tau_{\mathrm{sc,u}}$: a non-absorbing Rayleigh scattering layer above a semi-infinite, low albedo Rayleigh scattering atmosphere, the same scattering layer above a low albedo Lambertian surface, and an isotropic, non-polarizing scattering layer above the semi-infinite, low albedo Rayleigh scattering atmosphere.
    
The reflected polarization shows no dependence on the
polarization properties of the underlying surface for 
deep scattering layers with $\tau_{\mathrm{sc}}> 2$.
There are too many multiple scatterings to preserve this type of information from
deeper layers in the escaping photons. An imprint from the
polarization of the lower layer becomes visible 
for thin scattering layers with $\tau_{\mathrm{sc}}\lesssim 2$. Particularly 
well visible is the polarization dependence on $\tau_{\mathrm{sc}}$ for the
case where a polarizing layer is located below an isotropically
scattering layer. The polarizing lower layer only becomes apparent for $\tau_{\mathrm{iso}}<1$. 

The same is true for the polarized intensity, because the reflected intensity only shows
a very weak dependence on the phase function of the scattering layer. The effects are also very similar for the limb polarization at opposition.

\section{Wavelength dependence}
\label{sec:wavelength}

\begin{figure}
         \resizebox{\hsize}{!}{\includegraphics{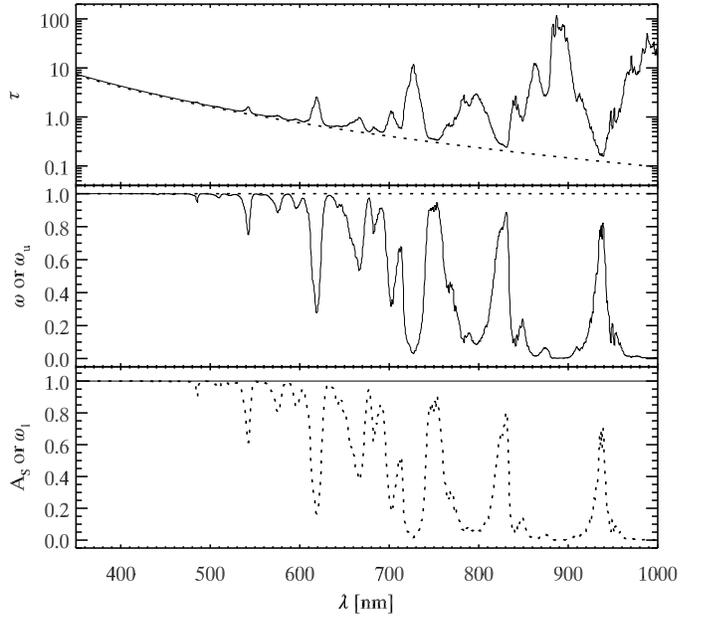}}
           \caption{Wavelength dependences of the model parameters total optical depth $\tau(\lambda) = \tau_{\mathrm{sc}} + \tau_{\mathrm{CH_4}}$ of the upper layer, single scattering albedo $\omega(\lambda)$ or $\omega_u(\lambda)$ of the upper layer, surface albedo $A_\mathrm{S}$ or single scattering albedo $\omega_l(\lambda)$ of the lower layer. Two cases are considered: A layer of Rayleigh scattering with CH$_4$ absorption above a white ($A_\mathrm{S}=1$) Lambert surface (solid), Rayleigh scattering layer without absorption $\omega_u=1$ above a deep clear atmosphere with methane absorption (dotted).}
                  \label{fig:lambda_params} 
\end{figure}

The wavelength dependence of the reflected intensity and polarization of a model
planet can be calculated using wavelength dependent parameters $\tau_{\mathrm{sc}}(\lambda)$,
$\omega(\lambda)$, and $A_\mathrm{S}(\lambda)$ or $\omega_\mathrm{l}(\lambda)$ for single or double layer models respectively. These parameters must be derived from a model with
a given column density of scattering particles and mixing ratios for   
Rayleigh scattering and absorbing particles. 

\begin{figure*}
         \resizebox{\hsize}{!}{\includegraphics{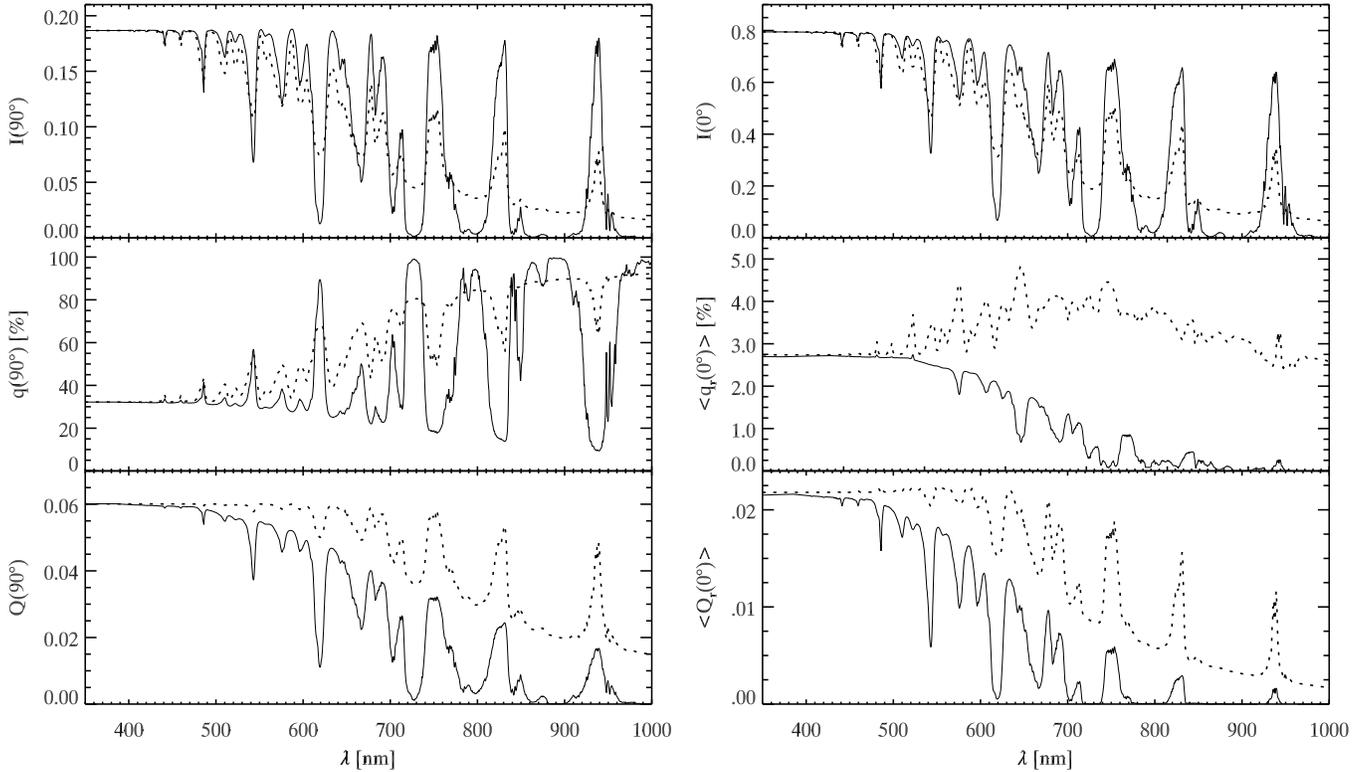}}
          \caption{Model spectra for the intensity, polarization, and polarized intensity
at quadrature (left) and intensity, radial polarization, and radial polarized intensity 
at opposition (right). Lines as in Figure \ref{fig:lambda_params}.
}
                  \label{fig:lambda_pol} 
\end{figure*}

As an example we selected parameters which approximate very roughly an Uranus-like atmosphere 
(e.g. Trafton~\cite{trafton76}) considering only Rayleigh scattering by H$_2$ and He and absorption
by CH$_4$. In our first example we look at a homogeneous scattering layer with methane absorption
above a reflecting cloud layer with a wavelength independent surface albedo $A_\mathrm{S}$ = 1. 
This is a strong simplification for Uranus because the methane mixing ratio is of order 100 lower in the stratosphere than in the troposphere (Sromovsky \& Fry \cite{sromovsky07}). Nevertheless it is a useful example for discussing basic effects of the wavelength dependence. 

In a second example we make a first approximation for a methane mixing ratio that is varying with height, by having an upper layer of finite thickness without methane and a lower semi-infinite layer that includes methane. 

The Rayleigh scattering cross section of molecular hydrogen is given by Dalgarno \& Williams (\cite{dalgarno62})
as 
\begin{equation}
\sigma_{\mathrm{Ray, H_2}}(\lambda) =  \frac{8.14 \cdot 10^{-13}} {\lambda^4} + \frac{1.28 \cdot 10^{-6}} {\lambda^6} + \frac {1.61} {\lambda^8}\,,
\end{equation}
where $\lambda$ is in \AA~and $\sigma_{Ray, H_2}$ in cm$^2$/molecule.

The total Rayleigh scattering optical depth is 
\begin{equation}
\tau_{\mathrm{Ray}} = \sigma_{\mathrm{ray, H_2}}\sum_i Z_i {(n_i-1)^2 \over (n_{H_2}-1)^2}\,,
\end{equation}
where $Z_i$ is the column density and $n_i$ the index of refraction of the $i$-th constituent\footnote{$n_{\mathrm{H_2}} = 1.0001384$, $n_{\mathrm{He}} = 1.000035$, $n_{\mathrm{CH_4}} = 1.000441$}. We use the same wavelength dependence as for the H$_2$ cross section for all constituents. 
Our upper scattering layer has a column density $Z=\sum_i Z_i$ = 500 km-am.\footnote{1 km-am = $2.687 \cdot
10^{24}$ molecules cm$^{-2}$} For the atmospheric composition we adopt particle fractions of 0.5\% CH$_4$ in the single layer case, and a methane free upper layer with 1 \% CH$_4$ in the lower layer in the two layer case. In all layers the He fraction is 15\% and the rest is H$_2$.

Because of the strong wavelength dependence of the Rayleigh scattering cross section, $\tau_{\mathrm{sc}}(\lambda)$ changes significantly from the UV to the near-IR (Fig.~\ref{fig:lambda_params}, top panel). Keeping $\omega$ and $A_\mathrm{S}$ fixed (no absorber)
yields the intensity and polarization results given in Figs.~\ref{fig:fp90_tau} and \ref{fig:fp0_tau} as function of $\tau_{\mathrm{sc}}$,  which are in this case equivalent to results as function of $\lambda$.

The wavelength dependent single scattering albedo $\omega(\lambda)$ follows from the CH$_4$ 
absorption optical depth $\tau_{\mathrm{CH_4}} = Z_{\mathrm{CH_4}} \kappa_{\mathrm{ CH_4}}(\lambda)$ and the Rayleigh scattering optical depth according to 
\begin{equation}
\omega(\lambda) = { \tau_{\mathrm{sc}}(\lambda) \over \tau_{\mathrm{sc}}(\lambda) + \tau_{\mathrm{CH_4}}(\lambda)}\,.  
\end{equation}
The absorption cross sections $\kappa_{\mathrm{CH_4}}(\lambda)$ were taken from Karkoschka (\cite{karkoschka94}) and the resulting $\omega(\lambda)$ is shown in Fig.~\ref{fig:lambda_params}. 

The intensity $I(\lambda)$, fractional polarization $q(\lambda)$, and polarized intensity $Q(\lambda)$
is determined from the wavelength dependent model parameters for our two cases at quadrature and opposition (Fig.~\ref{fig:lambda_pol}).

At quadrature both examples show similar results. In both cases the polarization is enhanced and the polarized intensity is reduced within methane absorption bands, only the changes are less pronounced for a non-absorbing upper layer. The polarized intensity $Q(\lambda)$ also drops with wavelength, but it is overall higher in the second case because the polarizing Rayleigh scattering extends to deeper layers. The biggest qualitative difference is seen in the continuum polarization $q(\lambda)$. In the case of an underlying reflecting cloud, $q(\lambda)$ drops towards longer wavelengths because of the smaller scattering optical depth above the diffusely scattering cloud. In the second case there is only polarizing Rayleigh scattering and no depolarization effect, so that the increasing absorption with wavelength in the lower layer results in a higher polarization. 

Similar spectropolarimetric models but with a Jupiter-like homogeneous atmosphere (higher column density, less methane than in our example) above both a dark surface $A_\mathrm{S} = 0$ and a reflecting extended cloud were discussed by Stam et al. (\cite{stam04}) for $\alpha = 90^\circ$. The qualitative behavior of intensity and polarization with wavelength is quite similar to our example. However, for the same column density and methane fraction we find a significantly lower intensity and higher polarization within methane bands. The origin of this discrepancy is unclear. A further comparison with intensity calculations for a Neptune-like atmosphere by Sromosvky (\cite{sromovsky05a}) shows a very good agreement at all wavelengths. Based on this we conclude that our model spectra should be correct.

Intensity and polarized intensity at opposition behave qualitatively similar to the large phase angle case. However the fractional polarization $q_r(\lambda)$ is completely different. Absorbing particles in the upper layer tend to reduce the fractional limb polarization, while absorption in the lower layer enhances it. Observations of the limb polarization of Uranus and Neptune (Joos \& Schmid \cite{joos07}) show that the fractional polarization is indeed enhanced within methane bands. Clearly for modeling limb polarization of these planets in absorption bands it is important to take into account the proper vertical stratification of the absorbing component. A detailed model accounting for methane saturation and freeze-out to fit the observations is beyond the scope of this paper.

\section{Special cases and diagnostic diagrams}
\label{sec:special}
We explore some special and extreme model cases in diagnostic diagrams
of observational parameters for phase angle $\alpha=90^{\circ}$ and opposition. 

\subsection{Fractional polarization versus intensity}

\begin{figure}
         \resizebox{\hsize}{!}{\includegraphics{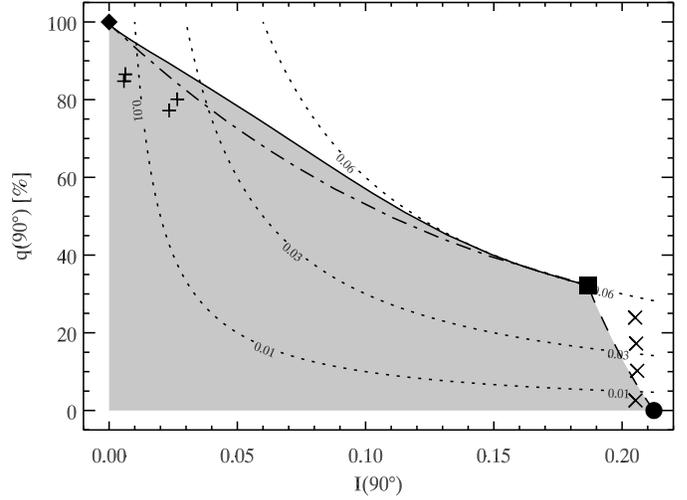}}
          \caption{Intensity vs. polarization at quadrature for the
            grid models. The shaded area indicates the possible range for Rayleigh and isotropic models. The symbols and lines indicate:
            semi-infinite conservative Rayleigh scattering (square), Lambert
            sphere (round), and black planet (diamond). The
            dash-dotted line shows semi-infinite models, the dashed
            and full line finite models without absorption ($\omega=1$) with surface albedo 
            $A_\mathrm{S}=1$ and 0 respectively. The haze models shown in Fig.~\ref{fig:phcurveshaze} are indicated by crosses (high albedo) and plusses (low albedo).}
          \label{fig:fp90_limits} 
\end{figure}

Figure~\ref{fig:fp90_limits} displays the diagnostic diagram for
the reflectivity $I(90^\circ)$ and the relative polarization $q(90^\circ)$ at phase angle
$\alpha=90^\circ$. Also indicated are the iso-contours for the
polarization flux $Q(90^\circ)$. 

\begin{figure}
         \resizebox{\hsize}{!}{\includegraphics{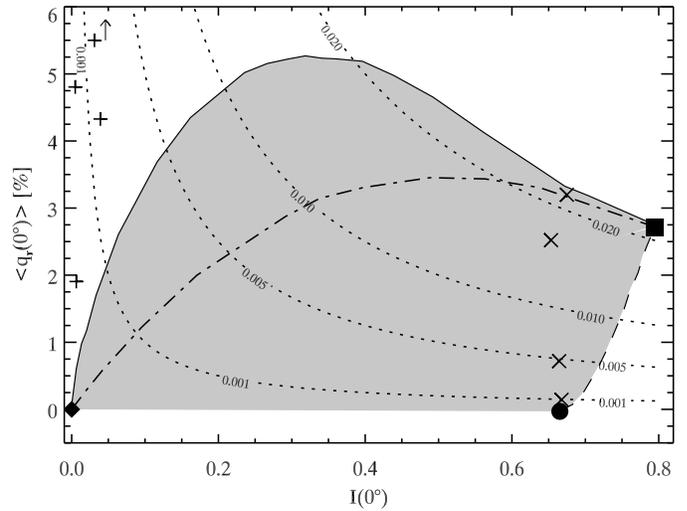}}
          \caption{Geometric albedo vs. disk-integrated radial
            polarization at opposition for the grid models. The
           models are the same as in Fig.~\ref{fig:fp90_limits}. 
%for the limiting cases of a finite atmosphere: $\omega=1$ and
%$A_\mathrm{S}=0$ (solid),1 (dashed) and for the semi-infinite atmosphere
%(dash-dotted). The shaded area indicates all possible results for our
% models.
} 
          \label{fig:fp0_limits} 
\end{figure}

The diagram shows points for special model cases and curves for
the dependence on specific model parameters. The shaded area 
defines the area of observational parameters covered by our 3-parameter model grid for Rayleigh scattering (Sect.~\ref{sec:homo}). Including isotropic scattering (Sect.~\ref{sec:isotropic}) or having a vertically inhomogeneous atmosphere (Sect.~\ref{sec:surface}) does not expand the covered area. 

Figure~\ref{fig:fp90_limits} emphasizes that it is not possible to have a Rayleigh scattering planet with both very high albedo and polarization. A high albedo implies either a lot of multiple or isotropic scattering, which both reduce the fractional polarization. On the other hand a high polarization implies mainly single scattering and therefore strong absorption and a low reflectivity. The maximal polarization at a fixed intensity is given by the model with a conservative $(\omega=1)$ scattering layer over a dark ($A_\mathrm{S}=0$) surface and appropriate $\tau_{\mathrm{sc}}$. The semi-infinite atmosphere with varying $\omega$ gives only slightly lower results than the former models. The maximum of the product $Q=q \cdot I=0.060$ is reached for the conservative semi-infinite atmosphere ($\tau_{\mathrm{sc}}=\infty$). Since the polarized intensity $Q$ is independent of the surface albedo $A_\mathrm{S}$, a change in $A_\mathrm{S}$ is equivalent to a shift along the $Q$~iso-contours in the diagram. 

The diagram also indicates the location of the haze models discussed in Sect.~\ref{sec:haze}. Most of the haze models lie within the same area as Rayleigh scattering. Only for very thick haze layers with high single-scattering albedo is it theoretically possible to get somewhat higher fractional polarization for a given intensity. 

Figure~\ref{fig:fp0_limits} is the same diagram at opposition
for the geometric albedo $I(0^\circ)$, the disk-integrated limb polarization $\langle q_r \rangle$
and iso-contours for the radial polarized intensity $\langle Q_r \rangle$. Like for large phase angles, the limb polarization for fixed intensity is highest for the conservative Rayleigh scattering layers over a dark surface. However, for $\tau_{\mathrm{sc}} \rightarrow 0$ the polarization drops to $0$\%, while at large phase angles with $A_\mathrm{S} = 0$ it raises towards $100$\% when the few reflected photons are mainly singly scattered. The semi-infinite models provide a distinctly lower fractional polarization signal than a finite conservatively scattering atmosphere over a dark surface. The fractional limb polarization for very low albedos can be significantly higher for atmospheres with haze than for Rayleigh scattering, because the unpolarized backscattering is greatly reduced. 

For models with two polarizing layers with different absorption properties, the results are located in the same area as for one layer above a surface. The limiting cases are models with a completely dark lower layer (equivalent to a dark surface), and two identical layers (equivalent to a single semi-infinite layer). For atmospheres that contain also isotropically scattering particles the polarization is always lower and the intensity either slightly enhanced or reduced depending on $\alpha$ because of the different scattering phase functions.

\subsection{Polarization near $\alpha=90^\circ$ versus limb polarization}

\begin{figure}
         \resizebox{\hsize}{!}{\includegraphics{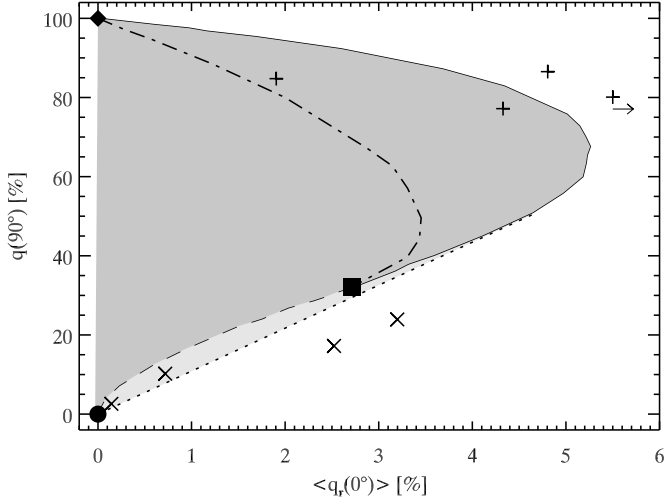}}
         \caption{Disk-integrated radial polarization at opposition
           vs. polarization at quadrature for the grid.
           The models are the same as in Fig.~\ref{fig:fp90_limits}. 
                   All Rayleigh scattering models lie in the dark shaded area, (partly) isotropic models also in the light shaded area. 
%limiting cases of a finite atmosphere: $\omega=1$ and $A_\mathrm{S}=0$
 %           (solid),1 (dashed) and for the semi-infinite atmosphere
  %          (dash-dotted). The light shaded area indicates all
   %         possible results for our Rayleigh scattering models, the
    %        dark shaded areas includes partly isotropic models.
} 
         \label{fig:p090_limits} 
\end{figure}

\begin{figure}
         \resizebox{\hsize}{!}{\includegraphics{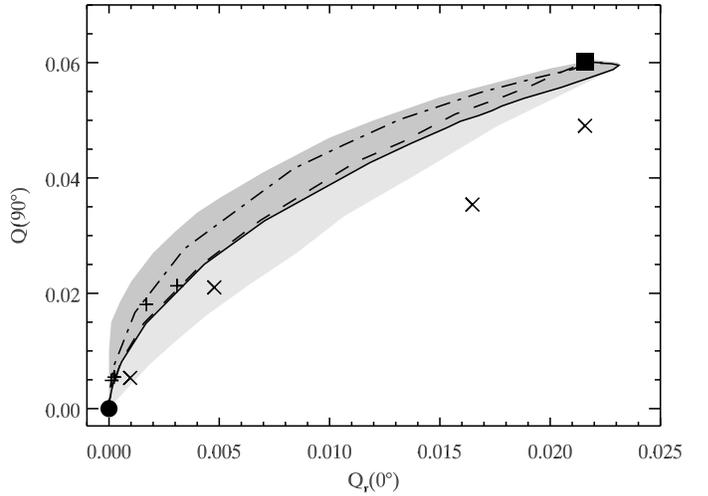}}
         \caption{Disk-integrated radial polarized flux at opposition
           vs. polarized flux at quadrature for the grid models.
The indicated models are the same as in Fig.~\ref{fig:fp90_limits}. All Rayleigh scattering models lie in the dark shaded area, (partly) isotropic models also in the light shaded area. 
% The dots are all results for Rayleigh scattering models, the crosses
% for partly isotropic models. A lower limit for the Rayleigh
% scattering models is given by the models with $A_\mathrm{S}=0,\omega=1$, for
% the partly isotropic models by $A_\mathrm{S} = 0, \omega_{\mathrm{iso}}+\omega_{\mathrm
%   Ray}=1$ and $\tau\sim2$
} 
         \label{fig:pf090_limits} 
\end{figure}

For the prediction or the future interpretation of the polarization of
extrasolar planets it is of interest to compare the polarization at phase angles near quadrature with the limb polarization at opposition. 

Figure \ref{fig:p090_limits} shows a diagram for the fractional
polarization $q(90^\circ)$ and the fractional limb
polarization $\langle q_r(0^\circ)\rangle$. Again the special models
are indicated with black symbols and lines as in 
Figs.~\ref{fig:fp90_limits} and \ref{fig:fp0_limits}. 

A lower limit for the polarization fraction at phase angles $\alpha \approx 90^\circ$ can be set from the limb polarization at opposition $\langle q_r \rangle$ for Rayleigh scattering or partly isotropic scattering atmospheres. For example a limb polarization of $\langle q_r \rangle \approx 2$\% 
implies a minimal polarization of $q(90^\circ)\approx20$\%. 
The upper limit for the polarization fraction $q(90^\circ$) is not 
well constrained by $\langle q_r \rangle$. The lower limit for Rayleigh or isotropic scattering may
overestimate the polarization at large phase angles only for very thick and bright haze layers.

A tighter correlation is obtained for the polarization flux
$Q(90^\circ)$ and the limb polarization flux $\langle Q_r \rangle$,
which is shown in Fig.~\ref{fig:pf090_limits}. All Rayleigh scattering models are located
in a narrow area along a line from the origin (Lambert sphere / black planet) to the semi-infinite, conservative Rayleigh scattering model. Thus, for Rayleigh scattering atmospheres, one can predict the large phase angle polarization flux $Q(\alpha)$ from the limb polarization flux $\langle Q_r
\rangle$ and vice versa. The area is slightly broadened if isotropic scattering is included in the models, but the relation still holds quite well. Only very thick and high albedo haze layers show a significantly lower $Q(90^\circ$) for a given $\langle Q_r(0^\circ)\rangle$. 

\subsection{Broadband polarized intensity}
\label{sec:color}

Color indices of observational parameters are often relatively easy to measure and they are helpful for the characterization of atmospheres. From the atmosphere models they are obtained by averaging spectral results (Sect.~\ref{sec:wavelength}) over the filter bandwidths.

\begin{figure}
         \resizebox{\hsize}{!}{\includegraphics{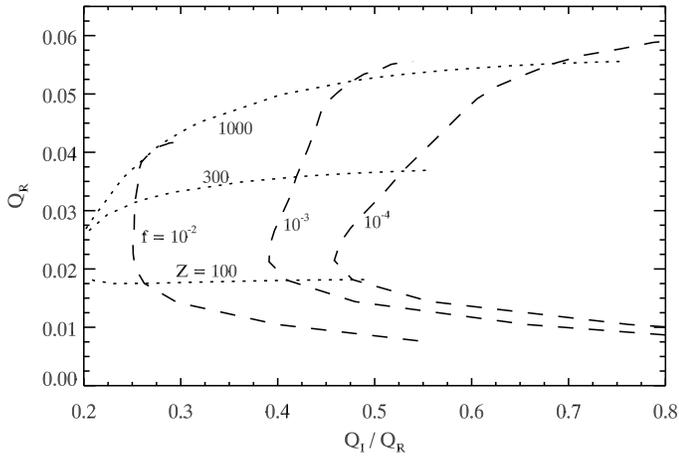}}
          \caption{Polarized intensity $Q_\mathrm{R}$ as a function of color $Q_\mathrm{I}/Q_\mathrm{R}$ where $Q_\mathrm{R}$ is the broad-band R signal (555 to 700 nm) and $Q_\mathrm{I}$ the broad-band I signal (715 to 865 nm) for Rayleigh scattering planets with methane at 90$^\circ$ phase angle. Indicated are models with constant methane mixing ratios $f$ (dashed) or constant atmosphere column density $Z$ in km-am (dotted) above a Lambert surface.}
 \label{fig:qintcolors} 
\end{figure}

Here we discuss the colors for a Rayleigh scattering atmosphere with methane as a main absorber. It is investigated how the polarized intensity color changes as a function of methane mixing ratio and column density above a cloud or surface. Color indices are calculated by integrating $Q(\lambda)$ over the wavelength range foreseen for filters in the {\sc sphere/zimpol} instrument (Beuzit et al. \cite{beuzit06}). The filters are assumed to have flat transmission curves with cut offs at 555 and 700 nm (R-band) and 715 and 865 nm (I-band). We concentrate on the color index of the polarized intensity $Q_\mathrm{I}/Q_\mathrm{R}$ (Fig.~\ref{fig:qintcolors}). 

The polarized intensity is higher at shorter wavelengths, and $Q_\mathrm{I}/Q_\mathrm{R} < 1$ for all models because of  the decrease of the Rayleigh scattering cross section with wavelength and the general increase of the absorption cross section of methane with wavelength. $Q_\mathrm{I}/Q_\mathrm{R}$ is near $1$ only for very thick atmospheres with very little methane or very thin atmospheres above a surface with wavelength independent scattering properties. In the former case $Q_\mathrm{R}$ and $Q_\mathrm{I}$ are very high, in the latter very low. For intermediately thick atmospheres the color index $Q_\mathrm{I}/Q_\mathrm{R}$ mainly depends on the methane mixing ratio, while $Q_\mathrm{R}$ mainly depends on the column density. 

From this diagram we may predict that a color index of $Q_\mathrm{I}/Q_\mathrm{R} \approx 0.25 - 0.5$ could be typical for Rayleigh scattering atmospheres with methane absorption. Aerosol particles and absorbers other than methane are expected to have a different spectral dependence of the scattering and absorption cross sections.

\section{Conclusions}

This paper presents a grid of model results for the intensity and polarization of 
Rayleigh scattering planetary atmospheres, covering the model parameter space in a systematic
way. The model parameters considered are the single scattering albedo
$\omega$, which describes absorption, the
scattering optical depth of the layer $\tau_{\mathrm{sc}}$, and the albedo
of a Lambert surface $A_\mathrm{S}$. The results of these model calculations 
are available in electronic form at CDS (see Appendix \ref{sec:cdstables}).
In addition we explore models which
combine Rayleigh and isotropic scattering, as well as particles with
strong forward scattering and atmospheres with vertical stratification.  

Simple Rayleigh scattering models are a good first approximation to
the polarization of light reflected from planetary atmospheres because
some amount of Rayleigh-like scattering by molecules or very small aerosol
particles can be expected in any atmosphere. From the model grid,
which basically provides monochromatic results, the
spectropolarimetric signal can be calculated. This is done by considering the
wavelength dependence of Rayleigh scattering and absorption in an
atmosphere with given column density and particle abundance (see Sect.~\ref{sec:wavelength}). 

The phase curves for the reflected intensity and polarization show a
strong dependence on phase but they always have similar shapes (see Fig.~\ref{fig:plotphaseradial}). However, the absolute level of the phase
curve is a strong function of atmospheric parameters such as the
abundance of absorbers or aerosol particles, the optical thickness of
the Rayleigh scattering layer, or the albedo of the surface layer
underneath (Sect.~\ref{sec:homo} and \ref{sec:beyond}, see also Kattwar \& Adams \cite{kattawar71}). 

The model calculations demonstrate that polarimetric observations
would provide strong constraints on the atmospheric properties of the
planetary atmospheres. An example is the polarization flux $Q(\alpha)$
of the reflected light which for optically thick atmospheres is a
simple function of the single scattering albedo roughly according to
$Q(\alpha)\propto \omega^b$ $(b\approx 1.5)$. If both polarization and
intensity can be measured, then one can distinguish between highly
reflective and absorbing planets with or without substantial layers of
Rayleigh-like scattering particles. 

According to the models a similar diagnostic is possible with
observations of the geometric albedo, center-to-limb polarization
profile, and limb polarization for solar system planets near
opposition. The limb polarization is in addition particularly
sensitive to the vertical stratification of scattering or absorbing
particles located high in the atmosphere. 

The diagnostic potential is further enhanced if data for different
spectral features, e.g. inside and outside of absorption bands, or
from different spectral wavelength regions can be combined (Sect.~\ref{sec:wavelength}, see also Stam et al.~\cite{stam04}). 

The calculations presented in this work are based on 
simple atmosphere models and they are therefore mainly useful for a
first interpretation of data. For spectropolarimetric data of high
quality, which are already available for solar system planets, one
should use more sophisticated atmospheric models including a more
detailed geometric structure, accurate abundances, and better
scattering models for aerosol particles. With such models it
might be possible for polarimetric studies to make a contribution to
our knowledge on the rather well known atmospheres of solar system objects.

Nonetheless the simple limb polarization models are of interest
because they link the model results for large phase angles, suitable
for extrasolar planet research, to models which can be easily compared
with observations of solar system objects. Thus it may be possible to
associate polarimetric observations of extrasolar planets to solar
system objects. On the other hand it is possible to predict the expected polarization
for quadrature phase of Uranus- and
Neptune-like extrasolar planets with this
simple model grid based on the existing limb
polarization measurements of Uranus and Neptune (Fig.~\ref{fig:pf090_limits}).  

Polarimetric measurements for extrasolar planets are expected in the
near future from high precision polarimeters. The measurements will first
provide the polarimetric contrast, which is the ratio of
the polarization flux from the planet $Q(\alpha)$ to the flux of the central star
according to 
\begin{equation} 
C(\alpha) = \frac{R^2}{D^2} \cdot q(\alpha) \cdot I(\alpha) = \frac{R^2}{D^2} \cdot Q(\alpha)\,,
\end{equation}
where $R$ is the radius of the planet and $D$ the distance from its
central star. 

Very sensitive polarimetric measurements of stars with known
close-in planets already exist, taken e.g. with the {\sc planetpol} instrument
(Hough et al.~\cite{hough06}; Lucas et al.~\cite{lucas09}). This instrument
measures the polarized intensity $Q(\alpha)$ of the planet diluted by the
unpolarized flux of the central star. It is then difficult to
separate the fractional polarization $q$ and the reflected intensity
$I$. $D$ is known from the radial velocity curve, but already the
radius of the planet $R$ may be hard to derive if the system shows no
transits. For photometrically very stable stars the reflected 
intensity $I(\alpha)\cdot R^2/D^2$ may be measurable with photometry of the phase 
curve with high precision instruments like {\sc most} (e.g. Rowe et al. \cite{rowe08}). 
An uncertainty in the planet radius will affect the precision of the
estimation of the normalized reflected intensity $I(\alpha)$ (or reflectivity) of the planet. 
 
{\sc sphere}, the future ``VLT planet finder'', which includes the 
high precision imaging polarimeter {\sc zimpol}, could provide
successful polarimetric detections ({Beuzit et al.~\cite{beuzit06}; Schmid et al.~\cite{schmid06b}}). This
instrument will be able to spatially resolve nearby ($d < 10$~ pc) star-planet systems and allow a polarimetric search for faint companions to stars. 
In a first step only the differential polarization signal,
i.e. the polarization flux $Q(\alpha)$, can be measured in the
residual halo of the central star. The measurement of the intensity
signal $I(\alpha)$ might be possible if further progress in 
techniques like angular differential imaging is achieved. Even if
a determination of $I(\alpha)\cdot R^2/D^2$ is possible, an
uncertainty remains in the translation to normalized intensity
$I(\alpha)$ if the radius of the planet is not known. 

Thus it may initially be quite difficult to measure intensity and
radius. For this reason it is important
to investigate the diagnostic potential of the wavelength
dependence in the polarization flux in more detail. For example the R-band and I-band 
yield a polarization color index $Q_\mathrm{I}/Q_\mathrm{R}$ from which it should be possible 
to infer constraints on the Rayleigh scattering optical depth or the 
strength of absorption bands (see Sect.~\ref{sec:color}). Another route of investigation for
atmospheres of extrasolar planets are measurements of the phase
dependence of the polarization flux. For example the location of the maximum of $Q$ 
along the phase curve is sensitive to the presence of aerosol particles, as discussed in Sect.~\ref{sec:haze}. 
For planets in eccentric orbits the dependence of the polarization flux on the separation from the
host star can be determined. 

One can hope that the current rapid progress in extrasolar planet observations continues, so that intensity measurements and accurate radius estimates for extrasolar planets
become available soon after the first polarization flux detections,
using the next generation of ground based telescopes and space
instruments. Such instruments, if equipped with a polarimetric observing mode, would allow a broad range of
observational programs on the reflected intensity and polarization from planets.

\begin{acknowledgements}
This work is supported by the Swiss National Science Foundation (SNSF). We thank Harry Nussbaumer and Franco Joos for carefully reading the manuscript. 
\end{acknowledgements}

\begin{appendix}

\section{Model grid tables}
\label{sec:cdstables}

\begin{table*}[htb!]
\caption{Extract of model grid results.}
\label{tab:cdstable}
\centering
\begin{tabular}{lllllllllllll}
\hline\hline
$\tau_{\mathrm{sc}}$ & $\omega$ & $A_\mathrm{S}$ & $A_{\mathrm{sp}}$ & $I(0^\circ)$ & $Q_r(0^\circ)$ & $I(7.5^\circ)$ & $I(12.5^\circ)$ & \ldots & $I(172.5^\circ)$  & $ Q(7.5^\circ)$ & \ldots & $ Q(172.5^\circ)$ \cr
\hline
30.00 &  1.00  & 1.0  &  1.0000 & 0.7947  & 0.02161   & 0.7846  & 0.7661  & \ldots & 0.0008 & 0.00334  & \ldots & -0.00001 \cr
99.00 &  0.99  & 1.0  &  0.7947 & 0.6378  & 0.02108   & 0.6290  & 0.6130  & \ldots & 0.0007 & 0.00331  & \ldots & -0.00001 \cr
99.00 &  0.95  & 1.0  &  0.5975 & 0.4884  & 0.01686   & 0.4813  & 0.4681  & \ldots & 0.0007 & 0.00304  & \ldots & -0.00001 \cr
99.00 &  0.90  & 1.0  &  0.4794 & 0.3980  & 0.01316   & 0.3918  & 0.3807  & \ldots & 0.0006 & 0.00278  & \ldots & -0.00001 \cr
99.00 &  0.80  & 1.0  &  0.3438 & 0.2912  & 0.00837   & 0.2866  & 0.2779  & \ldots & 0.0006 & 0.00229  & \ldots & -0.00001 \cr
99.00 &  0.60  & 1.0  &  0.1966 & 0.1707  & 0.00341   & 0.1676  & 0.1623  & \ldots & 0.0004 & 0.00146  & \ldots & -0.00000 \cr
99.00 &  0.40  & 1.0  &  0.1087 & 0.0958  & 0.00118   & 0.0940  & 0.0908  & \ldots & 0.0003 & 0.00084  & \ldots &  0.00000 \cr
99.00 &  0.20  & 1.0  &  0.0470 & 0.0418  & 0.00024   & 0.0410  & 0.0396  & \ldots & 0.0001 & 0.00039  & \ldots &  0.00000 \cr
99.00 &  0.10  & 1.0  &  0.0221 & 0.0197  & 0.00006   & 0.0193  & 0.0186  & \ldots & 0.0001 & 0.00018  & \ldots &  0.00000 \cr
10.00 &  1.00  & 1.0  &  1.0000 & 0.7949  & 0.02141   & 0.7848  & 0.7662  & \ldots & 0.0008 & 0.00338  & \ldots & -0.00001 \cr
10.00 &  1.00  & 0.3  &  0.8889 & 0.7085  & 0.02227   & 0.6992  & 0.6820  & \ldots & 0.0008 & 0.00338  & \ldots & -0.00002 \cr
10.00 &  1.00  & 0.0  &  0.8833 & 0.7042  & 0.02236   & 0.6950  & 0.6779  & \ldots & 0.0008 & 0.00336  & \ldots & -0.00002 \cr
10.00 &  0.99  & 1.0  &  0.8057 & 0.6453  & 0.02111   & 0.6378  & 0.6214  & \ldots & 0.0008 & 0.00310  & \ldots & -0.00001 \cr
10.00 &  0.99  & 0.3  &  0.7875 & 0.6326  & 0.02112   & 0.6238  & 0.6076  & \ldots & 0.0007 & 0.00342  & \ldots & -0.00003 \cr
10.00 &  0.99  & 0.0  &  0.7858 & 0.6312  & 0.02107   & 0.6225  & 0.6064  & \ldots & 0.0007 & 0.00344  & \ldots & -0.00002 \cr
\ldots & & & & & & & & & & & & \cr
\hline                                                                                                                                           
\end{tabular}                                                                                                                                                                                    
\end{table*}                                                                                                                                                                                     

{Our extensive model grid of intensity and polarization phase curves for homogeneous Rayleigh scattering atmospheres (Sect.~\ref{sec:homo}) is available in electronic form at CDS. Table \ref{tab:cdstable} shows a sample of the first few lines and columns. The table is structured as follows: Model parameters: {\it Column 1:} scattering optical thickness $\tau_{\mathrm{sc}}$, {\it Column 2:} single scattering albedo $\omega$, {\it Column 3:} surface albedo $A_\mathrm{S}$, Model results: {\it Column 4:} spherical albedo $A_{sp}$, {\it Column 5:} geometric albedo $I(0^\circ)$, {\it Column 6:} limb polarization flux $\langle Q_r(0^\circ)\rangle$, {\it Column 7:} $I(7.5^\circ)$, {\it Column 8:} $I(12.5^\circ)$,\ldots, {\it Column 40:} $I(172.5^\circ)$ {\it Column 41:} $Q(7.5^\circ)$,\ldots, {\it Column 74:} $Q(172.5^\circ)$. Columns 7 to 74 are $I(\alpha)$ and $Q(\alpha)$ spaced in 5 degree intervals. $I(0^\circ)$ is equivalent to $I(2.5^\circ)$ in our calculations. $Q(2.5^\circ)$, $I(177.5^\circ)$ and $Q(177.5^\circ)$ are very close to zero for all models and are not listed.

All results are disk-integrated. Binning, normalization and errors are described in Sect.~\ref{sec:monte}. For all calculations the number of photons was chosen such that $\Delta (Q/I) < 0.1 $\% for phase angles $\alpha = 0^\circ - 130^\circ$, and therefore $(\Delta I) / I < 0.07 $\% .

The model grid spans the following parameters: $\tau_{\mathrm{sc}} = 99$, 10, 5, 2, 1, 0.8, 0.6, 0.4, 0.3, 0.2, 0.1,  0.05, 0.01, $\omega = 1$, 0.99, 0.95, 0.9, 0.8, 0.6, 0.4, 0.2, 0.1, $A_\mathrm{S} = 1$, 0.3, 0. Models for only three values of $A_\mathrm{S}$ are given because the polarized intensity is independent of $A_\mathrm{S}$ and the intensity drops nearly linearly with increasing $A_\mathrm{S}$. Models with $\tau_{\mathrm{sc}} = 99$ were calculated only for $A_\mathrm{S}$ = 1 since the results are independent of $A_\mathrm{S}$. Instead of a model with  $\omega = 1$ and $\tau_{\mathrm{sc}} = 99$, we calculated the model with $\tau_{\mathrm{sc}} = 30$ to reduce computation time, but the results are equivalent. 

The spherical albedo $A_{\mathrm{sp}}$ in column 4 is the ratio of reflected photons in any direction to total incoming photons, while the geometric albedo $I(0^\circ)$ in column 5 is the disk-integrated reflected intensity at opposition normalized to the reflection of a white Lambertian disk. For our sample of Rayleigh scattering models, typically $I(0^\circ) = (0.80 \pm 0.06) A_{\mathrm{sp}}$. 
}

\end{appendix}

\end{document}